 \pdfoutput=1
\documentclass[10pt,letterpaper]{article}
\usepackage[top=0.85in,left=1.5in,footskip=0.75in]{geometry}

\usepackage{changepage}

\usepackage[utf8]{inputenc}

\usepackage{textcomp,marvosym}

\usepackage{fixltx2e}

\usepackage{amsmath,amssymb}

\usepackage{cite}
\usepackage{ulem}
\usepackage{float}

\usepackage{nameref}
\usepackage[hyperfootnotes=false]{hyperref}
\usepackage{siunitx}
\newcommand{\ms}{\si{\milli\second}}
\newcommand{\mum}{\si{\micro\meter}}

\usepackage[right]{lineno}

\usepackage{microtype}
\DisableLigatures[f]{encoding = *, family = * }

\usepackage{rotating}

\usepackage{color}

\usepackage{dsfont}
\usepackage[hang,bottom]{footmisc}

\usepackage[aboveskip=1pt,labelfont=bf,labelsep=period,justification=raggedright,singlelinecheck=off]{caption}

\makeatletter
\renewcommand{\@biblabel}[1]{\quad#1.}
\makeatother

\date{}

\usepackage{lastpage,fancyhdr,graphicx}
\usepackage{epstopdf}
\fancyheadoffset[L]{0.25in} 
\fancyfootoffset[L]{0.25in}

\newcommand{\xb}{\mathbf{x}}

\def\Suppl{Suppl. Inf. }
\def\SupplKpairwise{S1}
\def\SupplKpairwiseGibbs{S1.1}
\def\SupplKpairwiseGibbsBlockwise{S1.2}
\def\SupplFigConvergence{Suppl. Fig. S1}

\def\SupplFlatModelsHeat{S2.3}
\def\SupplFlatModelsHeatNonUnitTemp{S2.4}
\def\SupplFigNonNaturalSpikeCount{Suppl. Fig. S3}

\def\SupplCorrelationsWeak{S3.1}
\def\SupplCorrelationsBetaBinom{S3.2}

\begin{document}
\vspace*{0.35in}

\begin{flushleft}
{\Large
\textbf\newline{Signatures of criticality arise in simple neural population models with correlations}
}
\newline
\\
Marcel Nonnenmacher\textsuperscript{1,2,3*},
Christian Behrens\textsuperscript{3,4},
Philipp Berens\textsuperscript{3,4,5,6},
Matthias Bethge\textsuperscript{2,3,4},
Jakob H. Macke\textsuperscript{1,2,3*}
\\
\bigskip
\bf{1} research center caesar, an associate of the Max Planck Society, Bonn, Germany
\\
\bf{2} Max Planck Institute for Biological Cybernetics, T\"{u}bingen, Germany
\\
\bf{3} Bernstein Center for Computational Neuroscience, T\"{u}bingen
\\
\bf{4} Centre for Integrative Neuroscience and Institute of Theoretical Physics, University of T\"ubingen;
\\
\bf{5} Institute of Opthalmic Research, University of T\"{u}bingen
\\
\bf{6} Baylor College of Medicine, Houston, TX, USA
\bigskip

* marcel.nonnenmacher@caesar.de, jakob.macke@caesar.de

\end{flushleft}

\section*{Abstract}

Large-scale recordings of neuronal activity make it possible to gain insights into the collective activity of neural ensembles. It has been hypothesized that neural populations might be optimized to operate at a `thermodynamic critical point’, and that this property has implications for information processing. Support for this notion has come from a series of studies which identified statistical signatures of criticality in the ensemble activity of retinal ganglion cells. What are the underlying mechanisms that give rise to these observations?

Here we show that signatures of criticality arise even in simple feed-forward models of retinal population activity. In particular, they occur whenever neural population data exhibits correlations, and is randomly sub-sampled during data analysis. These results show that signatures of criticality are not necessarily indicative of an optimized coding strategy, and challenge the utility of analysis approaches based on equilibrium thermodynamics  for understanding partially observed biological systems.

\let\thefootnote\relax\footnote{Journal reference: Nonnenmacher M, Behrens C, Berens P, Bethge M, Macke JH (2017) \\Signatures of criticality arise from random subsampling in simple population models. \\PLoS Comput Biol 13(10): e1005718. https://doi.org/10.1371/journal.pcbi.1005718}

\section{Introduction}

Recent advances in neural recording technology \cite{Kerr_Denk_08,Marre_Amodei_12} and computational tools for describing neural population activity \cite{Tkacik_Marre_14} make it possible to empirically examine the statistics of large neural populations and search for principles underlying their collective dynamics \cite{Gao_Ganguli_15}. 
One intriguing hypothesis that has emerged from this approach is the idea that neural populations might be poised at a thermodynamic critical point \cite{ Beggs_Timme_12, Yu_Yang_13, Tkacik_Mora_15}, 
and that this might have important consequences for how neural populations process and encode sensory information
\cite{Tkacik_Mora_15, Mora_Deny_15}. As similar observations have been made in other biological systems  (e.g. \cite{Mora_Walczak_10,Bialek_Cavagna_12, Stephens_Mora_13}), it has been suggested that this might reflect a more general organizing principle \cite{Mora_Bialek_11}. 

In the case of neural coding, evidence in favour of this hypothesis has been put forward by a series of studies which measured neural activity from large populations of retinal ganglion cells and reported that their statistics resemble those of physical systems at a critical point \cite{Tkacik_Mora_15,  Mora_Deny_15}.
Using large-scale multielectrode array recordings \cite{Marre_Amodei_12}, spike-sorting methods that scale to large ($N>100$) populations \cite{Segev_Goodhouse_04, Marre_Amodei_12} and specially developed maximum entropy models \cite{Schneidman_Berry_06, Shlens_Field_06, Broderick_Dudik_07, Tkacik_Schneidman_09,  Ohiorhenuan_Mechler_10, Mora_Bialek_11, Tkacik_Marre_14,Roudi_Dunn_15}, 
Tka\v{c}ik et al. observed that the specific heat---a global population statistic which measures the range of probabilities of spike patterns---diverges as a function of population size. 
In addition, when an artificial temperature parameter is introduced, specific heat is maximised for the statistics of the observed data rather than for statistics which have been perturbed by changing the temperature parameter. 
These properties of retinal populations resemble the behaviour of physical systems at critical points, and gave rise to the hypothesis that neural systems might also be poised at critical points.

What neural mechanisms can explain these observations?  It had been hypothesised that the properties of the system need to be finely tuned \cite{Mora_Bialek_11, Tkacik_Mora_15} to keep the system at a critical point, for example through adaptation \cite{Shew_Clawson_15}.  A competing hypothesis \cite{Macke_Opper_11,Schwab_Nemenman_14,Aitchison_Corradi_14} had stated that generic mechanisms based on latent-variable models could be sufficient to give rise to activity data with these statistics, but neither of these theoretical studies had investigated mechanistic models of retinal population activity.
Thus, subsequent studies advocating criticality in the retina \cite{Tkacik_Mora_15,Mora_Deny_15}, continued to interpret their observations as indication for the retina to be poised at a special state that is advantageous for coding. It is therefore still an open question as to whether  previously reported signatures of criticality reveal a new mechanism of retinal coding, or  they are a direct consequence of the standard enconding models of retinal ganglion cell responses\cite{Chichilnisky_01, Carandini_Demb_05, Pillow_Shlens_08, Leen_Shea-Brown_15, Pitkow_Meister_12}. 

We here challenge the conclusion of studies which used tools from statistical physics to search for signatures of criticality by applying exactly the same data analysis approach to a simplistic feed-forward cascade model of retinal ganglion cell responses and showing that it exhibits the same effects.
Focusing on how the specific heat of this simulated data varies with population size and temperature, we show that this simple model exhibits signatures of criticality and reproduces the experimentally reported dependence on different stimulus ensembles \cite{Tkacik_Mora_15}. 
We provide a theoretical analysis of an analytically tractable model \cite{Amari_Nakahara_03, Macke_Opper_11,Tkacik_Marre_13}, and show mathematically that it exhibits signatures of criticality under a wide range of parameters.

This analysis also points to a subtle but important difference between how practical neural data analysis and theoretical studies often differ in how they study scaling behaviour of the system:
Whereas many theoretical studies describe different systems of size $N$, in practical neural data analysis populations of different size are typically constructed by randomly subsampling a large (but fixed) recording of neural activity.
We show that this sampling process produces 'signatures of criticality' whenever neural data has non-zero correlations, which could arise from a shared stimulus drive, recurrent connectivity or global state-fluctuations \cite{Harris_Thiele_11,Okun_Yger_12, Ecker_Berens_14, Scholvinck_Saleem_15}.

\section{Results}

\subsection{Signatures of criticality arise in a simple model of retinal ganglion cell activity}

A hallmark of criticality is that the specific heat of the model diverges when the temperature reaches the critical temperature \cite{Beggs_Timme_12}.
Tka\v{c}ik  et al.\ \cite{Tkacik_Mora_15} developed a statistical approach for translating this concept to neural data analysis. In their analysis, neural populations of different size $n$ are generated from the full recording by randomly subsampling the entire population.
The statistics of activity for each population of size $n$ are characterized using a maximum entropy model \cite{Schneidman_Berry_06, Shlens_Field_06, Tkacik_Schneidman_09,  Ohiorhenuan_Mechler_10, Tkacik_Marre_14}.  Finally, the maximum entropy models are perturbed by introducing a temperature parameter, and specific heat is computed for each population size $n$ and temperature $T$ from the (perturbed) maximum entropy model fit. Divergence of specific heat with population size $n$, and a peak of the specific heat near unit temperature $T=1$ (the 'temperature' of the original data)  are interpreted as evidence for the system being at a critical point \cite{Tkacik_Mora_15}.

\begin{figure}[H]
\vspace{-0.1cm}
\centering{
\includegraphics[width=0.77\textwidth]{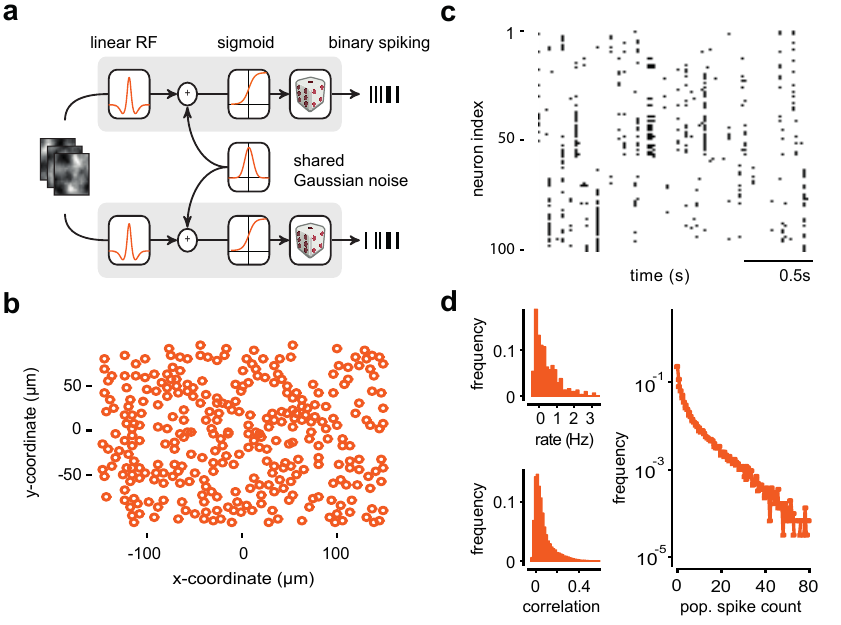} 	
\vspace{-0.1cm}
\caption{
{\bf A simple phenomenological model of retinal ganglion cell activity} 
{\bf a)} Model schematic: Neurons have linear stimulus selectivity with centre-surround receptive fields and also receive correlated Gaussian noise. Neural activity is modelled in discrete time. 
{\bf b)} Receptive field centres in simulation.
{\bf c)} Example raster plot of the simulated activity of $100$ neurons in response to natural stimuli.
{\bf d)} Statistics of population activity in response to natural stimuli. Histogram of firing rates (top left), correlation coefficients (bottom left) and frequency of population spike-counts (right). 
\label{fig_simulated}
}}
\end{figure}

To test if these signatures of criticality can  be reproduced by canonical properties of retinal circuits, we first created a simple phenomenological model of retinal ganglion cell (RGC) activity based on linear-nonlinear neuron models \cite{Chichilnisky_01, Carandini_Demb_05,Pitkow_Meister_12}. 
In this model (Fig. \ref{fig_simulated}a), we assumed retinal ganglion cells to have centre-surround receptive fields \cite{Kuffler_53, Pitkow_Meister_12} with linear spatial integration \cite{Rodieck_65}, sigmoid nonlinearities and stochastic binary spikes, i.e. in each time bin of size $20\ms$, each neuron $i$ either emitted a spike ($x_i=1$) or not $(x_i=0)$. 
We used a sequence of natural images (see Methods \ref{methods:simulations} for details).
In addition to the feed-forward drive by the stimulus, nearby neurons received shared Gaussian noise, mimicking common input from bipolar cells \cite{Trong_Rieke_08}. 
Thus, cross-neural correlations in the model arise from correlations in the stimulus, receptive-field overlap and shared noise, but not from lateral connections between RGCs. 
Parameters of the model were chosen to approximate the statistics of receptive-field centre locations of RGCs (Fig. \ref{fig_simulated}b), as well as histograms of firing rates, pairwise correlation-coefficients and population spike-counts (Fig. \ref{fig_simulated}d).
Nevertheless, the model clearly cannot accurately capture all statistics of real RGC activity: Our goal was not to provide a realistic model of retinal processing.
Rather, we wanted to directly test whether canonical mechanisms of retinal processing (overlapping centre-surround receptive fields, spiking nonlinearities, shared Gaussian noise) are sufficient for the signatures of criticality to arise, or whether this would require fine-tuning or sophisticated neural circuitry.

\begin{figure}[H]
\centering
\includegraphics[width=.9\textwidth]{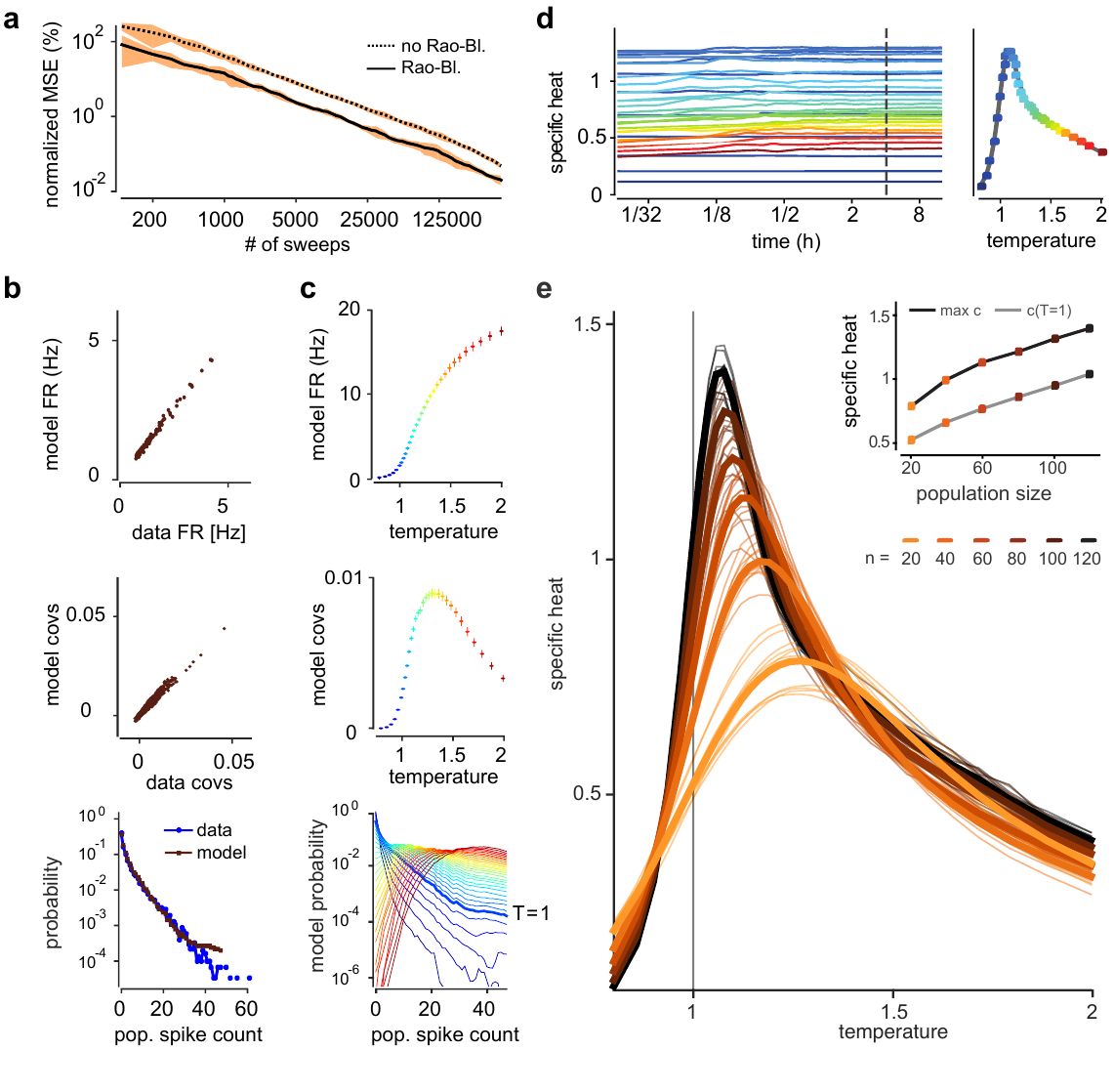} 	
\vspace{0.4cm}
\caption{
{\bf Signatures of criticality in a simple simulation of RGC activity}
{\bf a)}  Estimation-error (normalised mean square error) in pairwise covariances as function of sample size, averaged across $10$ populations of size $n=100$. Use of Rao-Blackwellization reduces the number of samples needed for a given level of accuracy by a factor of approximately $3$, making it possible to explore multiple large population models.
{\bf b)} Quality of fit: After convergence, the population models (here $n=100$, example population) capture the mean firing rates (top), covariances (centre) and spike count distribution (bottom) of the data.
{\bf c)} Changing the temperature parameter scales both mean firing rates (top), covariances (centre) and population spike-counts (bottom)
{\bf d)} Estimating specific heat via MCMC sampling: MCMC estimates of specific heat from a K-pairwise maximum entropy model fit to an example population (same model as in b-d). Final estimates were taken from the average over first $4 \si{\hour}$ sampling time. Right:  Resulting plot of specific heat as function of temperature. 
{\bf e)} Divergence of specific heat: Average and individual traces for $10$ randomly sampled populations for each of $6$ different population sizes, exhibiting divergence of specific heat and peak in heat near unit temperature. 
Inset: Specific heat at unit temperature and at peak vs. population size. 
\label{fig_criticality}}
\end{figure}

As a next step in the analysis, we subsampled populations of different size $n$ by uniformly sampling cells from our simulated recording of size $N=316$ neurons.
For each population we fit a 'K-pairwise' maximum entropy model  \cite{Tkacik_Marre_14}.
This model assigns a probability $P(\xb)$ to each spike-pattern $\xb$.  It is an extension of pairwise maximum entropy models (i.e. Ising models) \cite{Schneidman_Berry_06, Shlens_Field_06} which reproduces the firing rates and pairwise covariances which has additional terms which make sure that the model also captures the  population spike-counts of the data \cite{Tkacik_Marre_14} (see Fig. \ref{fig_simulated}d, and Methods \ref{methods:maxent} for details of model specification and parameterisation).
As we needed to efficiently fit this model \cite{Ferrenberg_Swendsen_88, Sohl_Battaglino_11,Schwartz_Macke_12} to multiple simulated data sets, we developed an improved fitting algorithm based on maximum-likelihood techniques using Markov chain Monte Carlo (MCMC) techniques (Matlab implementation available at https://github.com/mackelab/CorBinian), building on work by \cite{Broderick_Dudik_07}.
In particular, we made the most computationally expensive component of the algorithm, the estimation of pairwise covariances via MCMC sampling, more efficient by using a 'pairwise' Gibbs-sampling scheme with Rao-Blackwellisation \cite{Rao_45,Blackwell_47} (see Methods \ref{methods:maxent} for details). Rao-Blackwellisation resulted in a reduction of the number of samples (and computation time) needed for achieving low-variance estimates of the covariances by a factor of approximately  $3$ (Fig. \ref{fig_criticality}a,  \Suppl \SupplKpairwise). 
After parameter fitting, the model reproduced the statistics of the simulated data relevant for the model (Fig. \ref{fig_criticality}b).  
Using the formalism developed by Tka\v{c}ik et al., we then introduced a temperature parameter which rescales the probabilities of the model,
\begin{equation}
P_T(\xb)\propto P(\xb)^{1/T}.
\end{equation}

Here, temperature $T=1$ corresponds to the statistics of the empirical data. By changing $T$ to other parameter values one can perturb the statistics of the system \cite{Kirkpatrick_Gelatt_83}: 
Increasing temperature leads to models with higher firing rates and weaker correlations (Fig. \ref{fig_criticality}c), with $P_T(\xb)$ approaching the uniform distribution for very large $T$. 
If the  temperature is decreased towards zero, $P_T(\xb)$ has most of its probability mass over the most probable spike patterns. In many probabilistic systems,  lowering $T$ leads to  increasing correlations, as the systems then 'jumps' between several different patterns and thus the activation probabilities of different elements are strongly dependent on each other.
However, for the simulated RGC activity, the sparsity of data leads to a decrease of correlations:  At a bin size of $20$ $\ms$\cite{Schneidman_Berry_06}, the most probable state is the silent state, followed by patterns in which exactly one neurons spikes. In an example population of size $n=100$, $53.8\%$ of observed spike patterns contain at most one spike.
When decreasing the temperature to $T<1$, patterns with at most one spike dominate the systems even more strongly: For the same population and temperature $T=0.8$, we find $95.6\%$ of observed patterns to contain at most one spike. Thus when the temperature is lowered, the shift in probability mass to single-spike patterns decreases correlations.


We compute the specific heat of a population directly from the probabilistic model fit to data \cite{Tkacik_Mora_15}, using
\begin{align}
c(T) = \frac{1}{n} \mbox{Var}[\log P_T(X|\lambda)],
\label{eq:specific_heat}
\end{align}
i.e. the variance of the log-probabilities of the model, normalised by $n$ \cite{Tkacik_Mora_15}.
Specific heat is minimal for data in which all patterns $\xb$ that occur in the data are equally probable, and big for data in which pattern-probabilities span a  large range. We used MCMC-sampling to approximate the variance across all probabilities (see Methods \ref{methods:heat}), and used this approach to calculate, for each population of size $n$, the specific heat as a function of temperature  (Fig. \ref{fig_criticality}d). 

We found that the temperature curves obtained from the simulated data qualitatively reproduces the critical features of those that had been observed for large-scale recordings in the salamander \cite{Tkacik_Mora_15} and rat \cite{Mora_Deny_15} retina:  The peak of the curves diverges as the population size $n$ is increased, and moves closer to unit temperature for increasing $n$ (Fig. \ref{fig_criticality}e). 
Consistent with experimental findings, we found that specific heat diverged linearly with population size (Fig. \ref{fig_criticality}e). 
These results show that signatures of criticality arise in a simple feed-forward LN cascade model based on generic properties of retinal ganglion cells, and do not require finely tuned parameters or sophisticated circuitry. 



\subsection{Specific heat diverges linearly in flat population models}

\begin{figure}[H]
\centering
\includegraphics[width=1.0\textwidth]{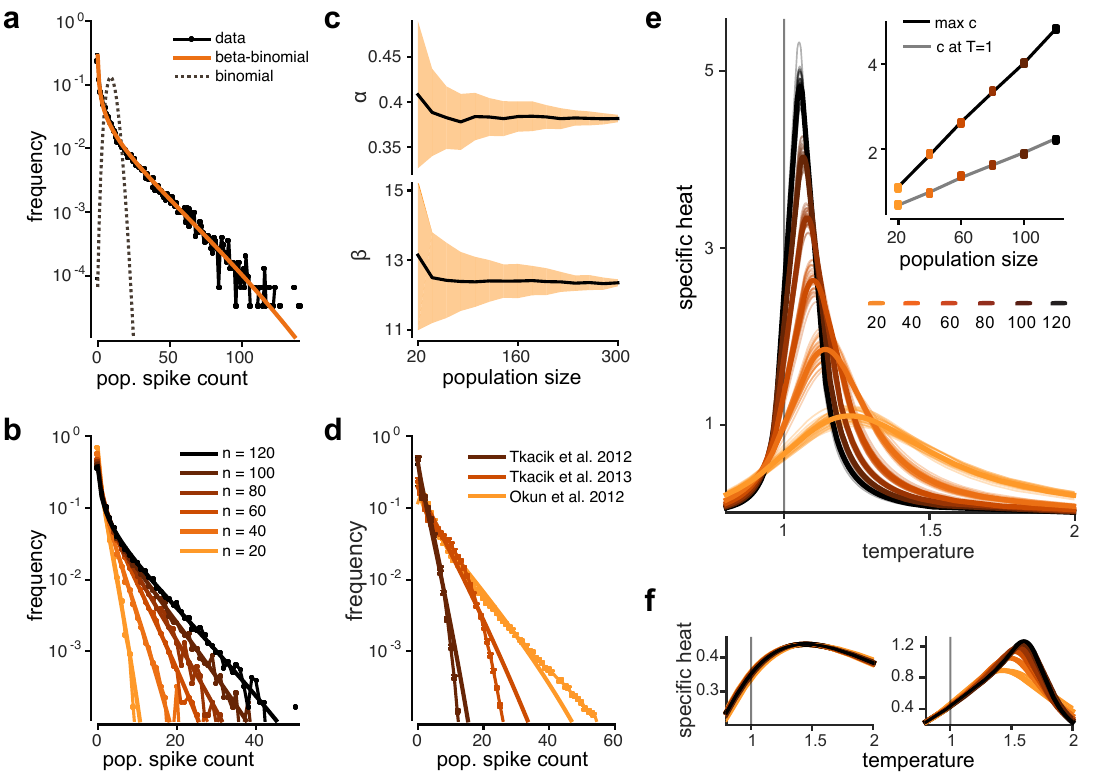} 	
\vspace{0.4cm}
\caption{
{\bf Signatures of criticality in a flat population model}
{\bf a)} Population spike-count distribution in RGC simulation,  and approximation by different models. Only the beta-binomial population model fits the simulated data accurately. 
{\bf b)} Beta-binomial model fits for different population sizes, indicating the goodness-of-fit is robust across population size.
{\bf c)} Estimates for beta-binomial parameters $\alpha$, $\beta$ for data from the simulation for different population sizes (mean $\pm$ 1 s.t.d.), Best-fitting parameters do not vary systematically with population size. 
{\bf d)} Beta-binomial model approximations to published empirically measured population spike-count distributions.
{\bf e)} Specific heat traces for the beta-binomial model, exhibiting signatures of criticality. Average and individual traces for $30$ randomly sampled populations for each of $6$ different population sizes. Inset: Specific heat at unit temperature and at peak vs. population size.
{\bf f)} Heat traces for independent model and flat pairwise maximum entropy model, which do not exhibit a divergence of the specific heat.
\label{fig_flat}}
\end{figure}

In the phenomenological population model above, we observed that specific heat grew linearly with population size, as it did in previous studies built on experimental data \cite{Tkacik_Schneidman_06,Tkacik_Mora_15,Mora_Deny_15}.  Can we understand this phenomenon analytically in a simplified model? 
In particular, is the divergence indeed linear, and what determines its rate? 
To address these questions, we replaced the K-pairwise maximum entropy model by a model which only captures the distribution of population spike-count $K = \sum_i x_i$ \cite{Macke_Opper_11, Tkacik_Marre_13, Amari_Nakahara_03,Okun_Yger_12} of the data, and in which all neurons have the same mean firing rate and pairwise correlations. 
This 'flat' model can be fit to data by matching its parameters to the population spike-count distribution, side-stepping the computational challenges of the K-pairwise model (see Methods \ref{methods:flat} for details). 
We  here introduce a new parametrised flat model in which the spike-count distribution is given by beta-binomial distribution $P(K | \alpha, \beta, n)$, reducing the number of free parameters from $n$ to $2$. The beta-binomial model is a straightforward extension of an independent (i.e. binomial) population model: At each time-point, a new firing probability $p$ is drawn from a beta-distribution with parameters $\alpha$ and $\beta$, and neurons then spike independently with probability $p$.  The fact that the underlying fluctuations in $p$ are shared across the population leads to correlations in neural activity.  This beta-binomial model provided a good fit to the population spike-count distributions of the simulated data (Fig. \ref{fig_flat}a) across different population sizes $n$ (Fig. \ref{fig_flat}b).
The best-fitting parameters $\alpha$ and $\beta$ did not vary systematically across population sizes, and converged  to values of $\alpha = 0.38$ and $\beta = 12.35$ (Fig. \ref{fig_flat} c), corresponding to an average firing rate of $\mu=1.5$ $\si{\Hz}$ (i.e. each neuron has a probability of spiking of $0.03$ in each bin) and average pairwise correlations of $\rho = 0.073$.
The beta-binomial model also provided good fits to population spike-count distributions published in \cite{Tkacik_Marre_13} and \cite{Okun_Yger_12}, \cite{Mora_Deny_15} (Fig. \ref{fig_flat}d). 
When we applied this flat model to populations subsampled from the RGC simulation, we could qualitatively reproduce the heat curves of the K-pairwise model. In particular, we  found a linearly diverging peak that moved closer to $T=1$ as the population size was increased (Fig. \ref{fig_flat} e).
Thus, linear divergence of specific heat is qualitatively captured by flat models.  We note that the absolute values of the specific heat do not match those of the K-pairwise model or simulated data, but are substantially bigger ($c_{max}= 4.02$ at $T=1.07$).

One of the difficulties of interpreting the scaling behaviour of maximum entropy models fit to neural data is the fact that the construction of the limit in $n$ differs from those studied in statistical physics: In statistical physics, different '$n$' typically correspond to systems of different total size, and the parameters are scaled as a deterministic function of $n$ (e.g. drawn from a Gaussian with variance proportional to $1/n$ in spin-glasses \cite{Sherrington_Kirkpatrick_75,Mezard_Parisi_87}).   In studies using maximum entropy models for neural data analysis, populations of different $n$ are obtained by randomly subsampling a fixed large recording, and the parameters are fit to each subpopulation individually. Thus, there is no analytical relationship between population size and parameter values in this approach, and this has made it hard to determine whether the scaling observed in these studies is surprising or not.

With the flat model, it is possible to analytically characterise the behaviour of the specific heat for large population sizes for this sampling process:  We assume that each population of size $n$ is randomly drawn from an underlying, infinitely large flat population model 
\cite{Macke_Opper_11, Amari_Nakahara_03}. 
Using this approach, one can mathematically show (see Methods \ref{methods:flat}, \Suppl \SupplFlatModelsHeat{}  and \cite{Macke_Opper_11} for details) that for virtually all flat models, the specific heat diverges linearly at unit temperature, but not for any other temperature $T>1$ or $T<1$ (\Suppl   \SupplFlatModelsHeatNonUnitTemp). 
As a consequence, the peak must move to $T=1$ as $n$ is increased.  Hence almost any data set analysed with the methods developed by \cite{Tkacik_Mora_15} will under the flat model exhibit signatures of criticality.
These results hold irrespective of the details of the properties of the full populations that the subpopulations are sampled from, including full populations that are more weakly or more strongly correlated than real neural populations, and even for models with unrealistic population spike-count distributions (see \SupplFigNonNaturalSpikeCount{} for an illustration).  
There are only two exceptions: The first one is a model in which all neurons are independent (i.e. a binomial population model), and the second one is a flat pairwise maximum entropy model---indeed, this is the only flat model with non-vanishing correlations for which the specific heat does not have its peak at unit temperature (see Fig. \ref{fig_flat}f and \cite{Macke_Opper_11} for an illustration).

\subsection{Strong neural correlations lead to fast divergence of specific heat.}

The rate at which the specific heat diverges provides a mean of quantifying the 'strength' of criticality. What is the relationship between correlations in a neural population and the rate of divergence?
To study how the specific heat rate $\tilde c=c(T=1)/n$ depends on the strength of correlations, we used a beta-binomial model to generate simulated data with firing rate $\mu=1.5$ $\si{\Hz}$ (i.e. each neuron has a probability of spiking of $0.03$ in each bin), and different (population-wide, as all neuron pairs have the same correlation) pairwise correlation coefficients $\rho$ ranging from $\rho=0.01$ to $\rho=0.25$ (Fig. \ref{fig_correlations}a). 
We found  that the heat curves had the same shape as in the analyses above, with a peak that increases and moves to unit temperature (Fig. \ref{fig_correlations}b).
Comparing the results for different specified correlation strengths within the populations, we found that the specific heat rates $\tilde c$ increased strictly monotonically with $\rho$ (Fig. \ref{fig_correlations}b,c). 
For the beta-binomial model, the large-n value of $\tilde c$ can be calculated analytically (see \Suppl  \SupplCorrelationsBetaBinom{} for details) as a function of the parameters  $\alpha$ and $\beta$,
\begin{align} 
\tilde c &= \frac{\alpha (\alpha +1) \psi_1(\alpha+1) + \beta (\beta +1) \psi_1(\beta+1)}{ (\alpha + \beta) (\alpha + \beta +1 ) } \nonumber\\
  &+ \frac{\alpha \beta \left( \psi_0(\alpha +1) - \psi_0(\beta +1) \right)^2 }{ (\alpha + \beta)^2 (\alpha + \beta +1 ) } - \psi_1(\alpha + \beta +1 )       \label{eq:tildec_betabin}.
\end{align}

This analytical evaluation of $\tilde c$ (valid for large $n$) was in good agreement with numerical simulations (Fig. \ref{fig_correlations}c left). 
In the case of weak correlations $\rho$, equation \ref{eq:tildec_betabin} can be simplified: In this case, the specific heat rate is proportional to the strength of correlations (see \Suppl  \SupplCorrelationsWeak{} for details), i.e. 
\begin{equation}
\tilde c \approx \rho \mbox{ } \mu(1-\mu) \left(\log\left(\frac{1-\mu}{\mu}\right) \right)^2 \label{eq:tildec_general}
\end{equation}
This expression can also be derived from the Gaussian model in \cite{Mora_Deny_15} equation (4), by inserting the expected values of the mean and variance of the  population spike-count under random subsampling. Thus, at least for flat models and the analysis based on specific heat proposed previously,  'being very critical' is a consequence of 'being strongly correlated'.  

\subsection{Specific heat depends on average correlation strength in K-pairwise model}
\label{results:stimulusTypes}

Is the relationship between the strength and correlations and the 'strength' of criticality (i.e. the divergence rate of specific heat) also true in more general models? In the original study \cite{Tkacik_Mora_15}, specific heat was computed from K-pairwise model fits to RGC activity resulting from three different kind of stimuli:
Checker-board stimuli (which do not have long-range spatial correlations, although stimulus-driven cross-neural correlations can arise from receptive field overlap), natural images, which exhibit strong spatial correlations, and full-field flicker (which constitutes an extreme case of spatial correlations since all pixels in the display are identical). 
Tka\v{c}ik et al. found that specific heat diverges in all three conditions, and interpreted this as evidence that signatures of criticality are not `inherited from the stimulus' \cite{Tkacik_Mora_15}.
Comparing the specific heat values for $n=100$ reported in \cite{Tkacik_Mora_15} across stimulus conditions, Tka\v{c}ik et al. found the smallest peak for checkerboard stimuli ($c_{max}  = 0.54$ for $n=100$), intermediate for natural images ($c_{max} = 0.92$) and strongest for full-field flicker ($c_{max}  = 2.4$). 

\begin{figure}[H]
\centering
\includegraphics[width=.8\textwidth]{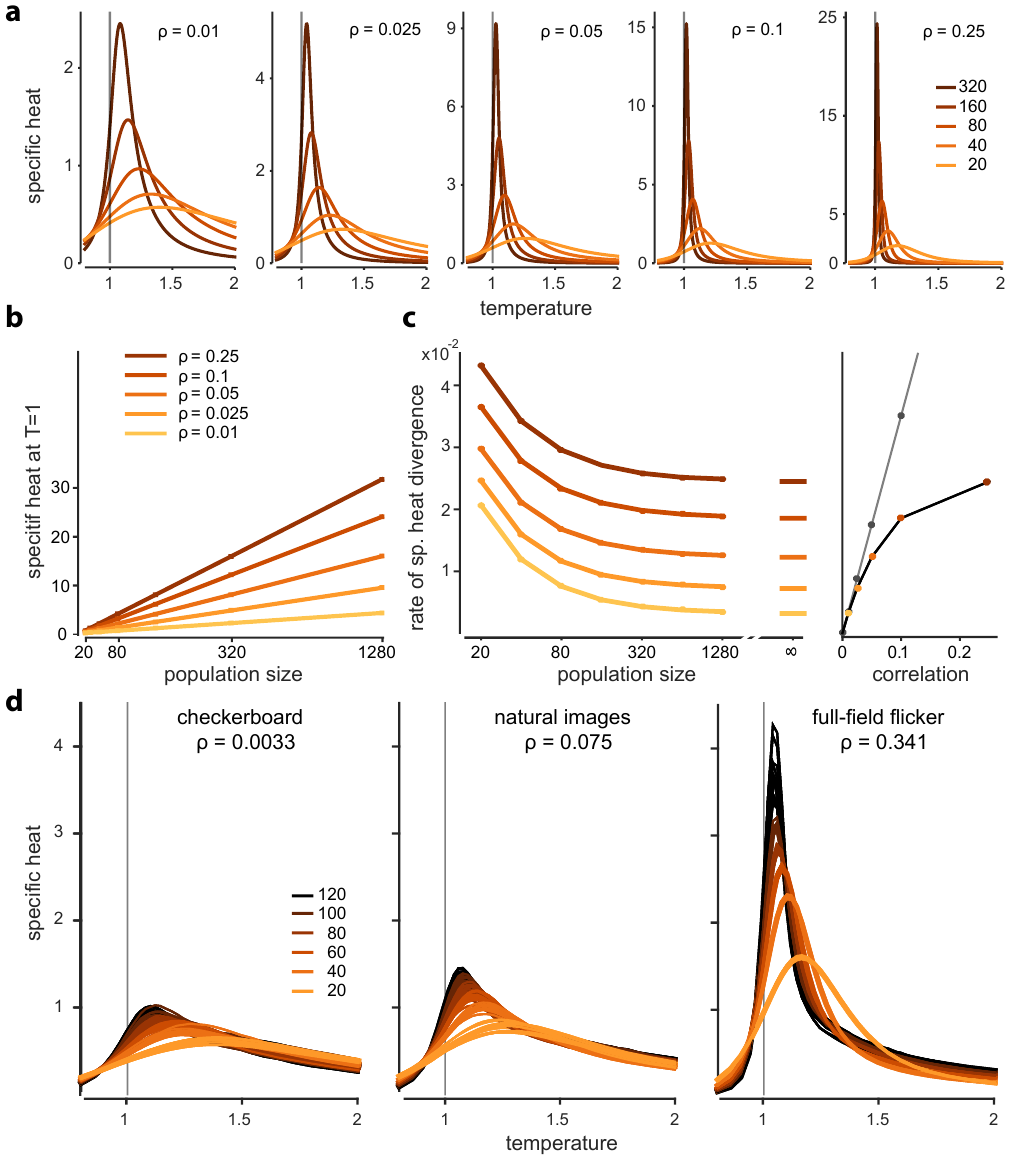} 	
\vspace{0.4cm}
\caption{
{\bf Relationship between correlations and criticality.}
{\bf a)} Specific heat traces for beta-binomial model of different correlation strengths and population sizes. Heat traces are qualitatively similar, but differ markedly quantitatively (see y-axes).
{\bf b)} Specific heat diverges linearly, and the slope depends on the strength of correlations (left). 
Divergence rate of specific heat for beta-binomial model as a function of correlation strength (centre). Rightmost point (at infinity) corresponds to analytical prediction of large-$n$ behaviour.  Divergence rates are strictly increasing with correlation strength (right) which is captured by a weak-correlation approximation (dashed line). 
{\bf c)} Specific heat increases with correlation  in the K-pairwise maximum entropy model: average and individual traces for $10$ randomly subsampled populations for $6$ different population sizes. Left to right: checkerboard, natural images and full-field flicker stimuli presented to the population. Correlation strengths denote mean correlation coefficient in each population.  
\label{fig_correlations}}
\end{figure}

We tested whether we find the same pattern of results in K-pairwise model fits to our retinal simulation. Specific heat divergence also followed the pattern predicted by the flat models (Fig. \ref{fig_correlations}d):  
Checkerboard (which gave an average correlation between neural activity of  $\rho=0.033$) had the smallest peak  (peak specific heat $c_{max} = 0.87$) followed by natural images ($\rho=0.075$, $c_{max} = 1.32$) and full-field flicker ($\rho=0.341$, $c_{max} = 3.09$).
We conclude that the experimental evidence---which showed that the specific heat diverges, and how the speed of divergences depends on the stimulus ensemble---is entirely consistent with a simple, feed-forward phenomenological model of retinal processing.  

\subsection{Sources of criticality-inducing correlations in neural activity}

\begin{figure}[H]
\centering
\includegraphics[width=.8\textwidth]{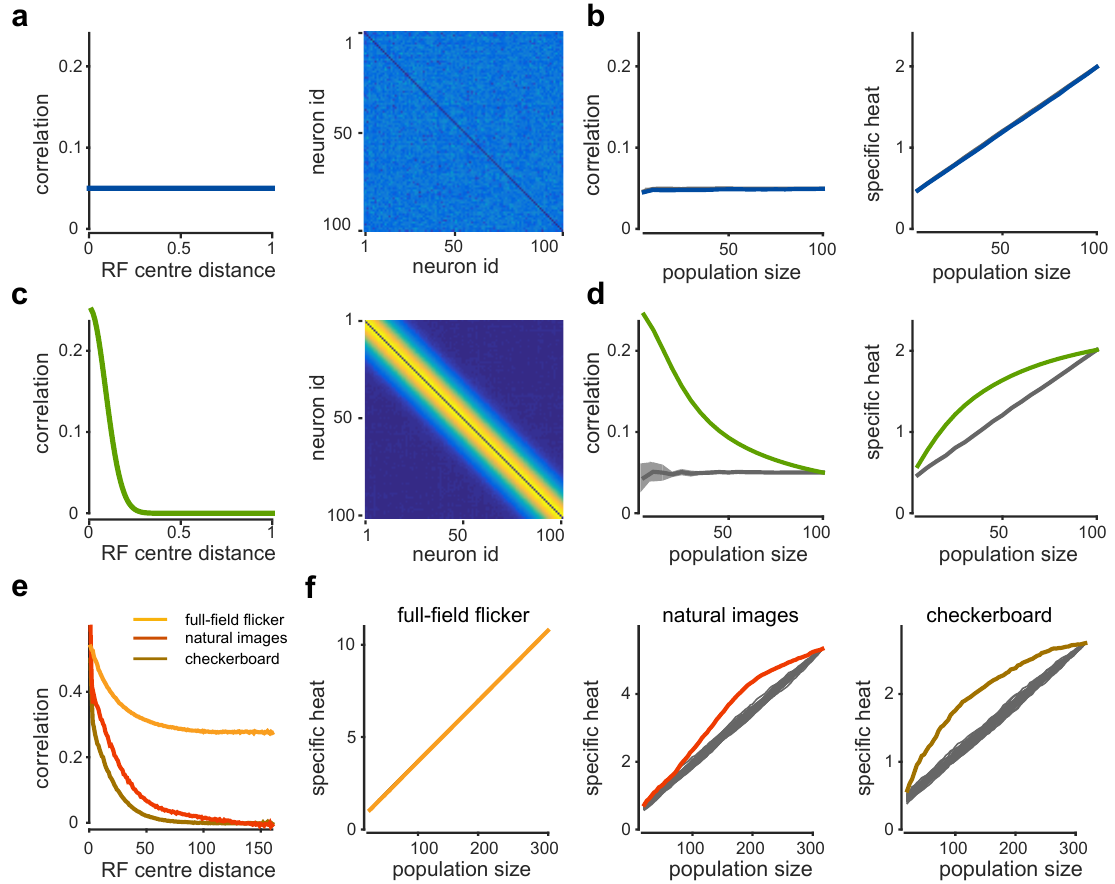} 	
\vspace{0.4cm}
\caption{
{\bf Random subsampling leads to criticality-inducing correlations.} 
{\bf a)} Illustration: A population with 100 neurons and infinite-range correlations, the average correlation between any pair of neurons is close to $0.05$. Correlation as function of inter-neuron distance (left) and full correlation matrix (right).
{\bf b)} Average correlation in subpopulation of different size $n$ (left) and specific heat as function of $n$ (right), when neurons are sampled from $1$ to $100$. Random sampling gives identical results (not shown).  
{\bf c)} Population with limited-range correlations, same plots as in panel a. 
{\bf d)} Left: Average correlation as function of population size for ordered sampling (green) and uniform subsampling (gray). Right: Specific heat grows linearly for random subsampling, but shows signs of saturation for ordered sampling.
{\bf e)} Average correlation as function of inter-neuron distance in RGC simulation. For checkerboard and natural images, correlations drop to $0$ for large distances. 
{\bf f)} Specific heat for different stimulation conditions, for ordered (colour) or random subsampling (gray).  
\label{fig_sampling}
}

\end{figure}

In the above, we showed that a beta-binomial spike count distribution can be sufficient for signatures of criticality to arise. For this to hold 
we need the variance of the population spike-count to grow at least quadratically with population size, i.e. $\mbox{Var}(K) \propto n^2$.  The variance of the population spike-count is equal to the sum of all variances and covariances in the population, $\mbox{Var}(K)= \sum_{i=1}^n \mbox{Var}(x_i)+ \sum_{i\neq j} \mbox{Cov}(x_i,x_j)$. 
A sufficient condition for signatures of criticality to arise in these models is that the average covariances (and hence  correlations) between neurons are independent of $n$, $\frac{1}{n(n-1)} \sum_{i \neq j} \mbox{Cov}(x_i,x_j) \approx \mbox{constant}$ \cite{Yu_Yang_13, Beggs_Timme_12}.  One possible correlation structure which has this properties are so called 'infinite range' correlations (Fig. \ref{fig_sampling}a): correlation between  neurons do not drop off to $0$ for large spatial distances. In this case, adding more and more neurons to a population will not change the average pairwise correlation within the population (Fig. \ref{fig_sampling}b). 

In neural systems, there are at least two reasons that can facilitate the required correlation structure. First, as shown above, the choice of stimuli has a clear effect on the heat capacity indicating an important effect of input-induced correlations. In particular for full-field flicker stimuli infinite-range correlations are to be expected but also white noise input can generate correlations of considerable extent due to overlapping receptive fields. Second, even a neural population which does not have infinite range correlations can appear critical if it is randomly subsampled during analysis: Suppose that different populations of size $n$ are obtained as above by (uniformly) subsampling a large recording of size $N$. Then, for any correlation structure on the full recording (including limited-range correlations, Fig. \ref{fig_sampling}c), the average correlation in a population of size $n$ will be independent of $n$  (Fig. \ref{fig_sampling}c):  If neurons are randomly subsampled from the large recording, then the pairwise correlations in each subpopulation are also a random subsample of the large correlation matrix.
As a consequence, the average correlation will be independent of $n$, and specific heat will diverge with constant slope (Fig. \ref{fig_sampling}d). In contrast, if different population sizes are constructed by taking into account the spatial structure of the population (i.e. by iteratively adding neighbouring cells) then the average correlation in each subpopulation will drop with $n$, and the slope of specific heat growth will decrease with population size.

In our RGC simulation, correlations did drop off to zero with spatial distance for checkerboard and natural images, but not for full-field flicker (Fig. \ref{fig_sampling}e). Correlations in the full-field flicker condition initially drop off due to distance-dependent shared noise, but eventually saturate at a level far above zero that is determined by the full-field stimulus.
Due to these strong infinite-range correlations, both spatially structured sampling and uniform sampling then give rise to linear growth in specific heat (Fig. \ref{fig_sampling}f left).
For the other two stimulus conditions, however, the choice of subsampling scheme does result in markedly different behavior of the specific heat growth: Both for natural images and checkerboard stimuli, we can see the rate of growth decreases for large $n$ under spatially structured subsampling (Fig. \ref{fig_sampling}f centre, right). This effect will be more pronounced for larger simulations, and in additional simulations we found specific heat to saturate completely once populations are substantially bigger than the spatial range of correlations. 

In summary, populations will exhibit critical behaviour if correlations have infinite range (over the size of the recording), irrespective of the sampling scheme. In addition, if a population is randomly subsampled (as was done in \cite{Tkacik_Mora_15,Mora_Deny_15}), then signatures of criticality will arise even if the underlying correlations have limited range.

\section{Materials and Methods}

\subsection{Numerical simulation of retinal ganglion cell activity}
\label{methods:simulations}

\paragraph{Retina simulation:} We simulated a population of $N = 316$ retinal ganglion cells as linear threshold neurons whose receptive fields were modelled by difference-of-Gaussian filters with ON-centres \cite{Rodieck_65, Pitkow_Meister_12, Carandini_Demb_05}. The simulation comprised two subgroups of cells with different receptive field sizes (surrounds $56 \mum$ and $30 \mum$ in retinal space, centres $28 \mum$ and $15 \mum$, respectively, one third cells with large receptive fields).
For both subgroups, the weight of the surround was $0.5$ of the centre weight. 
Locations of receptive field centres were based on a reconstruction of $518$ soma locations from a patch of mouse retina \cite{Baden_Berens_15}. As the reconstructed locations in that data set also comprised about $40\%$ amacrine cell somata, we randomly discarded  $40\%$ of the cell locations.
The resulting patch of retina covered an area of $200 \times 300 \mum^2$, corresponding to $100 \times 150$ pixels in stimulus space. 
Correlated noise across neurons was modelled using correlated additive Gaussian noise. Correlations dropped off exponentially with soma distance with a  decay constant of $\tau=30 \mum$ i.e. noise covariance matrix was chosen as $\Sigma=\sigma_{noise}^2(a I_n+b e^{-\Delta/\tau})$, where $\Delta_{ij}$ is the distance between neurons $i$ and $j$ and $a^2+b^2=1$. We set $\sigma_{noise}=0.022$ and $a=0.45$. 
We modelled neural spiking in discrete time using $20\ms$ bins. In each bin $t$, the total input $z_i(t)$ to neuron $i$ was given by $z_i(t)= w_i^\top s(t)+ \epsilon_i(t)$, where $w_i$ is the receptive field of neuron $i$, $s(t)$ the vectorised stimulus and $\epsilon_i(t)$ the input noise of neuron $i$.
A neuron in a given bin is active ($x_i = 1$) if  $z_i + d > 0.5$ and inactive ($x_i = 0$) otherwise, with offset $d=0.168$.
Parameters of the simulation (centre and surround sizes, relative strength of centre and surround, magnitude and correlations of noise, spiking threshold) were chosen to roughly match the statistics of neural spiking (firing rates, pairwise correlations, population activity counts) reported in studies of salamander retinal ganglion cells \cite{Schneidman_Berry_06, Tkacik_Marre_14, Marre_Amodei_12}. (Code will be available at www.mackelab.org/code). 


\paragraph{Stimuli:} We used three types of stimuli for this study: natural images, checkerboard patterns and full-field flicker. For natural image stimuli, we used a sequence of $101$ images of meadow sceneries taken from low hight. Each image was $400 \times 400$ pixels, and each image was presented for $20\ms$ with 300 repetitions total. The luminance histograms of the images were transformed to a normal distribution with mean 0.5 and pixel values between 0 and 1.  

For the full-field flicker stimulus, luminance levels were drawn from a Gaussian distribution with mean $\mu=0.5$ and variance $\sigma^2=0.06$. Checkerboard stimuli consisted of $80 \times 80$ tiles of size $5 \times 5$ pixels each.
Luminance levels (from within the interval $[0,1]$) of each tile were chosen to be either $0.15$ or $0.77$ with probability $0.5$. The parameters of both  stimulus sets were chosen to match the dynamic range of the simulated retinal ganglion cells. For both types of stimuli, $2000$ images were generated and the image sequences were presented with $10$ repetitions. 
To calculate specific heat as function of increasing population size, we randomly selected $10$ subsamples of the full simulated population of $N = 316$ cells at population sizes $n \in \{20, 40, 60, 80, 100, 120\}$ by uniformly drawing $n$ neurons out of the full population without replacement.

\subsection{Modeling neural population data with maximum entropy models}
\label{methods:maxent}

\paragraph{Model definition:} We modelled retinal ganglion cell activity by using a 'K-pairwise' maximum entropy model \cite{Tkacik_Marre_14}. In a maximum entropy model \cite{Jaynes_57}, the probability of observing the binary spike word $\xb \in \{0, 1\}^n$ for parameters $\lambda=\{h,J,V\}$ is given by 
\begin{align}
P(\xb|\lambda) = \frac{1}{Z(\lambda)} \exp\left( h^\top \xb + \xb^\top J \xb +  \sum_{k=0}^n  V_k \delta\left( K(\xb)=k \right) \right)
\label{eq:maxent}
\end{align}
Here, the parameter vector $h$ (of size $n \times 1$) and the upper-triangular matrix $J \in \mathbb{R}^{n \times n}$ correspond to the bias-terms and interaction terms in a pairwise maximum entropy model (also known as an Ising model or spin-glass) \cite{Schneidman_Berry_06}. The term $K(\xb)=\sum_{i=1}^n x_i$ denotes the population spike-count, i.e. the total number of spikes across the population within a single time bin, and the indicator-term $\delta\left(K=k\right)$ is $1$ whenever the population spike-count equals $k$, and is $0$ otherwise.
The term $\sum_{k=0}^n  V_k \delta\left(K=k\right)$ was introduced by \cite{Tkacik_Marre_14} to ensure that the model precisely captures the population spike-count distribution of the data using $n$ additional free parameters. The partition function $Z$ for given $\lambda$ is chosen such that the probabilities of the model sum to $1$.

\paragraph{Parameter fitting:} To fit the model parameters $\lambda=\{h, J,V\}$ to a data set $D=\{\xb^{(1)}, \xb^{(2)}, \ldots, \xb^{(M)}\}$, we maximised the penalised log-likelihood \cite{Dudik_Schapire_06, Altun_Smola_06} of the data $D$ under the model, 
\begin{align}
L(h,J,V):&=\sum_{m=1}^M \log P(\xb^{(m)} | h,J,V)  - \frac{1}{\sigma_h}\|h\|_1 - \frac{1}{\sigma_J} \|J\|_1 - \frac{1}{2} V^T \Sigma^{-1} V \label{eq:penloglikelihood}. 
\end{align}
Here, the $l1$-penalty controlled the magnitudes of parameters $h$, $J$, the term $\|J\|_1$ favoured sparse coupling matrices, and the regularisation term $\Sigma$ on the $V$-parameters ensures that the terms controlling the spike count distribution vary smoothly in $k$ (\Suppl  \SupplKpairwise). This smoothness prior is particularly important for large spike counts, as it makes it possible to interpolate parameters  for which the number of observed counts is small.

In maximum entropy models, exact evaluation of the penalised log-likelihood and its gradients requires the calculation of expectations under the model, $\text{E}[x_i]$, $\text{E}[x_i x_j]$ or equivalently $\mbox{cov}(x_i, x_j)$, and $P(K=k)$ (\Suppl  \SupplKpairwiseGibbs), which in turn requires summations over all $2^n$ possible states $\xb$ and is prohibitive for $n>20$.
Following previous work \cite{Broderick_Dudik_07}, we used Gibbs sampling to approximate the relevant expectations (\Suppl  \SupplKpairwiseGibbs{} for derivations and implementation details). We used two modifications over previous applications of Gibbs sampling to fitting maximum entropy models to neural population spike train data, with the goals of speeding up parameter learning and alleviating memory usage: 

First, we use Rao-Blackwellisation \cite{Rao_45, Blackwell_47} to speed up convergence of the estimation of  covariances of $\xb$. We used pairwise Gibbs sampling (blocked Gibbs with block size $2$), where each new sample in the MCMC chain was obtained by updating two entries $i$ and $j$ of $\xb$ at a time, rather than just a single entry.
This allowed us to get estimates of the conditional probabilities $P( x_i x_j=1 | x_{\sim \{i,j\}})$, and to use them to speed up the estimation of the second moment $\mbox{E}[x_i x_j]$ from empirical average of these conditional probabilities (\Suppl  \SupplKpairwiseGibbs). 

Second, we used a variant of coordinate ascent that calculated all relevant quantities as running averages over the MCMC sample, and thereby avoided having to store the entire $n \times \tilde{M}$ MCMC sample in memory \cite{Broderick_Dudik_07}, where $\tilde{M}$ is the length of the sample. Because all features of the maximum entropy model are either $0$ or $1$ ($x_i$, $x_ix_j$ and the indicator function for the spike count), the gain in log-likelihood obtainable from either updating a single element of $h$ or $J$ \cite{Broderick_Dudik_07,Schwartz_Macke_12}, or from updating all $V$ simultaneously (but not from updating multiple entries of $h$ and $J$) can be computed directly from MCMC estimates of $\text{E}[x_i]$, $\text{E}[x_i x_j]$ and $P(K=k)$ (\Suppl  \SupplKpairwiseGibbsBlockwise).
For each iteration, we calculated the gain in log-likelihood for each possible update of $h_i$, $J_{ij}$ and full $V$, and picked the update which led to the largest gain \cite{Dudik_Phillips_04,Broderick_Dudik_07}. 

We measured the length of Markov chains in sweeps, where one sweep corresponds to one round of $n(n-1)/2$ Markov chain updates that encompasses all pairs of entries of $\xb$ in random order. We set a learning schedule that started at $800$ sweeps for the first parameter update and doubled the number of sweeps in the chain after each set of $1000$ parameter updates.
We monitored convergence of the algorithm using a normalised mean square error between empirical $\text{E}[x_i]$, $\mbox{cov}(x_i, x_j)$, $P(K=k)$ and their estimates from the MCMC sample. For normalisation, we used the average squared values of the target quantity, e.g. $\frac{1}{n} \sum_{i=1}^n <x_i^2>$ for the firing rates. We stopped the algorithm when a pre-set threshold was reached ($0.01\%$, $0.25\%$, $0.01\%$ for $\mbox{E}[x_i]$, $\mbox{cov}(x_i, x_j)$, $P(K=k)$, respectively), or when the fitting algorithm took more than $\left( \frac{n}{100} \right)^2 \times 72 \si{\hour}$ of computation time on a single core ($2.294$ GHz AMD Opteron(TM) Processor 6276)  (\SupplFigConvergence). For $10$ populations of size $n=100$ (for natural images), the normalised MSEs after model-fitting were $0.43\%$, $2.80\%$, $0.42\%$).
An implementation of the fitting algorithms in MATLAB is available at https://github.com/mackelab/CorBinian. 




\subsection{Calculating specific heat and temperature curves}\label{methods:heat}

\paragraph{Specific heat calculations:} To investigate thermodynamic properties of neural population codes, Tka\v{c}ik et al \cite{Tkacik_Mora_15} introduced a temperature parameter $T$ for equation \ref{eq:maxent}: 
\begin{equation}
P_T\left( \xb |  \lambda \right) = \frac{1}{Z_T} \exp\left( \frac{1}{T} \left( h^\top \xb + \xb^\top J \xb +  \sum_{k=0}^n  V_k \delta\left( K(\xb)=k \right) \right) \right)
\label{eq:maxentwithT}
\end{equation}

Model fits are obtained at $T=1$, and the temperature parameter $T$ is scaled to study the system (i.e. characterised by  $P_T\left( \xb |  h, J, V \right)$ for $T~=1$). We note that varying $T$, in effect, modulates probabilities by exponentiating them with $1/T$,
\begin{equation}
P_T(\xb) \propto \left(P_{T=1}(\xb)\right)^{1/T},
\end{equation}
and that the family of probability distributions obtained by varying $T$ can be constructed for any distribution, not just maximum entropy models. For large temperatures $P_T$ approaches a uniform distribution ($P_T(\xb) \approx 2^{-n}$ for each $\xb$), whereas for small temperatures it converges to a singleton, $P_T(\xb^*)\approx 1$ with $\xb^*=\mbox{argmax}_{\xb}(P_{T=1}(\xb))$.

The specific heat, as given in equation \ref{eq:specific_heat}, can be obtained from the variance of the log-probabilities of the model.
As the variance in practice can not be outright computed for $n$ beyond $20$, we obtained estimates of $c(T)$ using a pairwise Gibbs sampler. We note that the specific heat does not depend on $Z_T$, as changing $Z_T$ results in a constant, additive shift in log-probabilities which does not affect the variance.  We tracked the variance of log-probabilities over an MCMC chain $\xb^{(1)}, \ldots, \xb^{(\tilde{M})}$ of length $\tilde{M}$ sampled at temperature $T$,
\begin{equation}
c(T) \approx \frac{1}{n} \left( \frac{1}{\tilde{M}} \sum_{m=1}^{\tilde{M}} \left(\log P_T\left( \xb^{(m)} |  \lambda \right)\right)^2  - \left( \frac{1}{\tilde{M}} \sum_{m=1}^{\tilde{M}} \log P_T\left( \xb^{(m)} |  \lambda \right) \right)^2  \right).
\end{equation}
For each population, we evaluated $c(T)$ for $31$ temperatures between $T=0.8$ and $T=2$, and found the Gibbs sampler to provide reliable estimates over this temperature range. 
We used a burn-in of $2*10^4$ sweeps, and ran the sampler for  $\left(\frac{n}{100}\right)^2 \times 4 \si{\hour}$ of CPU time, resulting in between $9.97$e$5$ and $1.72$e$6$ sweeps (mean $\pm$ std) for $n=100$ (i.e. between $4.94$e$9$ and $8.52$e$9$ sampled individual spike words).

\subsection{Simplified population models and the beta-binomial model}
\label{methods:flat}


For the theoretical analysis, we adopted a class of population models (here referred to as 'flat' models) in which all neurons have identical mean firing rates, pairwise correlations and higher-order correlations \cite{Amari_Nakahara_03,  Macke_Opper_11, Yu_Yang_11, Tkacik_Marre_14, Barreiro_Gjorgjieva_14}. 
Such a model is fully specified by the population spike-count distribution $P(K=k)$, and all spike words with the same spike count are equally probable. As a result, the probabilities of individual patterns $\xb$ can be read off from the spike count distribution by
\begin{equation} 
P(\xb)={n \choose k}^{-1} P(K=k)
\end{equation}
whenever $\sum_{i=1}^n x_i = k$. In a maximum entropy formalism, this model can be obtained by setting $h_i=0$ and $J_{ij}=0$ for all $i,j \in \{1, \ldots, n\}$ and only optimising entries of $V$. Without loss of generality, we fixed fixed $V_0 = 0$ \cite{Tkacik_Marre_13}, resulting in $n$ degrees of freedom for the model.  

In flat models, it is possible to explicitly construct a limit $n \rightarrow \infty$ which will help us understand population analyses performed on experimental data: We assume that there is a spike count density $f(r)$, $r \in [0,1]$, which describes the population spike-count distribution of an infinitely large population. $f(r)$ denotes the probability density of a fraction of $r$ neurons spiking simultaneously. Finite-size populations of $n$ cells are then obtained as random subsamples out of this infinitely large system. Based on previous findings by \cite{Macke_Opper_11}, we show in \Suppl  \SupplFlatModelsHeat{} that, in this construction, flat models always exhibit a linear divergence of specific heat, unless the limit $f(r)$ is given by either a single delta peak or a mixture of two symmetric delta peaks.
These two models corresponds to systems that (for large $n$) either behave like a fully independent population (whose spike count distribution converges to a single delta peak), or a population described by a pure pairwise maximum entropy model (which converges to two delta peaks). In particular, any flat model with higher-order correlations \cite{Ohiorhenuan_Mechler_10,Yu_Yang_11,Barreiro_Gjorgjieva_14,Leen_Shea-Brown_15}, or a non-degenerate $f(r)$, will exhibit 'signatures of criticality'. Furthermore, we show that, for continuous $f(r)$, $c(T)$ does not diverge for any $T\neq 1$. In combination, these results show that the peak of the specific heat is mathematically bound to converge to $T=1$ for $n \rightarrow \infty$ in this model class.



We further simplified the flat model by re-parametrising $P(K=k)$ by a beta-binomial distribution, thereby reducing the number of parameters from $n$ to two, and---importantly---obtaining parameters which do not explicitly depend on $n$. In this model, 
\begin{equation}
P(K=k)={n \choose k} \frac{\mbox{Beta}(\alpha + k, \beta + n - k)}{\mbox{Beta}(\alpha, \beta)}
\end{equation}
and 
\begin{equation}
f(r)= \frac{1}{\mbox{Beta}(\alpha,\beta)}r^{\alpha-1} (1-r)^{\beta-1}. 
\end{equation}
For simulated data, we found values for $\alpha$, $\beta$ extracted from the beta-binomial fits to populations of different sizes $n$ to be stable over a large range of $n$ (Fig. \ref{fig_flat}b). 
We used the beta-binomial parameters obtained from the largest investigated $n$ to estimate the divergence rate $\tilde c$ for $n \rightarrow \infty$.  

\section{Discussion}

An intriguing hypothesis about the collective activity of large neural populations has been the idea that their statistics resemble those of physical systems at a critical point. 
Using a definition of criticality which is based on temporal dynamics with power-law statistics, numerous studies have reported and studied critical behaviour in neural population activity \cite{Beggs_Plenz_03, Petermann_Thiagarajan_09,Mora_Deny_15,Shew_Clawson_15}.  Multiple possible mechanisms for these dynamics have been proposed (e.g. \cite{Levina_Herrmann_07,Shew_Clawson_15,Markram_Muller_15}). It has been argued that such temporal dynamics might be beneficial for neural computation and communication \cite{Bertschinger_Natschlaeger_04,Shew_Yang_11,Shew_Clawson_15} (see \cite{Beggs_Timme_12} for an overview). 
More recently, a second line of studies  \cite{Mora_Bialek_11, Beggs_Timme_12, Yu_Yang_13, Stephens_Mora_13,Tkacik_Mora_15, Mora_Deny_15} has studied the statistics of time-instantenous patterns of neural activity using tools from statistical mechanics, and argued that they also exhibit critical behaviour.
This hypothesis could open up further questions on how the system maintains its critical state, and what implications this observation has for how neural populations encode sensory information and perform computations on it. Similarly, signatures of criticality have also been observed in natural images \cite{Stephens_Mora_13} and small cortical populations \cite{Yu_Yang_13}, and have been studied using  the theory of finite-size scaling and critical exponents \cite{Yu_Yang_13}.  It has been argued that systems close to a critical point might be optimally sensitive to external perturbations \cite{Yu_Yang_13} and that the large dynamic range of the code (i.e. the large variance of log-probabilities) might be beneficial for encoding sensory events which likewise have a large distribution of occurrence-probabilities \cite{Tkacik_Schneidman_09}. 

Alternatively, generic mechanisms could be sufficient to give rise to activity data with these statistics. We  had demonstrated in a previous theoretical study \cite{Macke_Opper_11} that a simple models with common input can exhibit signatures of criticality. More recently, Schwab et al. \cite{Schwab_Nemenman_14} and Aitchison et al. \cite{Aitchison_Corradi_14} elaborated on these findings, showing that common input (or other latent variables which lead to shared modulations in firing rates) can give rise to Zipf-like scaling of pattern probabilities (a second signature of criticality).    Mathematically, Zipf's Law is equivalent to stating that the plot of entropy vs energy (i.e. log-probability) is a straight line with unit slope \cite{Schwab_Nemenman_14, Aitchison_Corradi_14}. Schwab et al \cite{Schwab_Nemenman_14} showed that particular latent variable models give rise to Zipf's law.
This result was generalized by \cite{Aitchison_Corradi_14} which showed that,  under fairly general circumstances, high-dimensional latent variable models exhibit a wide distribution of energies (i.e. log-probabilities) and hence a large specific heat. In addition, they showed that large fluctuations in the specific heat are (under some additional assumptions) sufficient to achieve Zipfian scaling. While it has also been argued that the use of data-sets which are too small might give rise to spuriously big specific heats \cite{Saremi_Sejnowski_14}-- while this is true in principle, additional analyses e.g. in \cite{Tkacik_Mora_15} show that their results are robust with respect to data-set size.

However, neither of these previous theoretical studies analysed mechanistic models of neural population activity, nor did they have  tools for studying population statistics in large simulations or recordings, and  they were therefore limited to studying very small ($N<20$) systems.
It has thus been an open question of whether and how these theoretical considerations can account for effects  observed in retinal ganglion cells.  We here showed that surprisingly simple mechanisms are sufficient for two key signatures of thermodynamic criticality---a divergence of specific heat and a peak of the specific heat near unit temperature--- to arise.  

We found that neural population activity exhibits signatures of criticality whenever the average correlation in population of different sizes is larger than zero and does not depend on population size.  In the thermodynamic analysis of physical systems at equilibrium, long-range correlations typically vanish in the thermodynamic limit. In neural systems, however, such 'criticality-inducing' correlations can arise as a consequence of various factors: In a local patch of retina, retinal ganglion cells have a large degree of receptive field overlap, and natural stimuli also contain strong spatial correlations.
This can lead to correlations which do have unlimited range within the experimentally accessible length scales. Thus, fluctuations in the stimulus will lead to common activity modulations amongst neurons within the population. Empirically, activity correlations between pairs of retinal ganglion cells only fall of slowly with the distance between somas (or receptive field centres) \cite{Pitkow_Meister_12}. 
Similarly, Mora et al \cite{Mora_Deny_15} used a moving-bar stimulus with strong temporal correlations, and found that including activity from multiple time-lags markedly increase the strength of specific heat. We hypothesise that this increase in specific heat is a consequence of temporal correlations being stronger than inter-neural correlations in this stimulus condition.
In addition, firing rates of cortical neurons are modulated by global fluctuations in excitability \cite{Harris_Thiele_11,Okun_Yger_12,Ecker_Berens_14,Scholvinck_Saleem_15}, resulting in neural correlations with infinite range.  

Finally, we showed that criticality-inducing correlations arise as a consequence of constructing  different subpopulations by uniformly subsampling a large recording with correlations. Signatures of criticality are entirely consistent with canonical properties of neural population activity, and require neither finely-tuned parameters in the population, nor sophisticated circuitry or active mechanisms for keeping the system at the critical point.  Signatures of criticality are likely going to be found not just in retinal ganglion cells, but in multiple brain areas and model systems.  These observation raise the question of whether signatures of criticality are really indicative of an underlying principle, or rather are a consequence of viewing the statistics of neural populations through the lens of equilibrium thermodynamics. In order to realise the potential of large-scale recordings of neural activity in the search of a theory of neural computation, we will need data-analysis methods which are adapted to the specific properties of biological data \cite{Gao_Ganguli_15,Roudi_Dunn_15}. 

\section{Acknowledgments}
Work was funded by the German Federal Ministry of Education and Research (BMBF; FKZ: 01GQ1002, Bernstein Center T\"ubingen, FKZ 01GQ1601 to PB), the German Research Foundation (BE 5601/1-1 to PB; SFB 1233, Robust Vision: Inference Principles and Neural Mechanisms, TP 14 to JHM), the Max Planck Society and the caesar foundation. 
We thank F. Franzen for help with figures and cluster computing, S. Buchholz, D. Greenberg  and S. Turaga for comments on the manuscript and useful discussions. 
The funders had no role in study design, data collection and analysis, decision to publish, or preparation of the manuscript.

\nolinenumbers

\bibliography{critical_retina}


\renewcommand{\thesection}{S\arabic{section}}
\renewcommand{\thefigure}{S\arabic{figure}}
\setcounter{figure}{0}
\setcounter{section}{0}
\setcounter{subsection}{0}

\clearpage

\clearpage

\centerline{\huge{Supporting Information}}
\vspace{0.5cm}




\section{Fitting the K-pairwise maximum entropy model to data}
\label{fitting:overview}

To identify the values $\hat{\lambda}$ of the model parameters which yield the best fit of the maximum entropy model to data, we maximise the log-likelihood of the model given the data. The general form of the log-likelihood of a maximum entropy model parametrised by vector $\lambda$ is given by
\begin{align}
L(\lambda) = \sum_{m=1}^M \log P(\xb^{(m)}| \lambda)= - M \log {Z_\lambda}  + \sum_{m=1}^M \lambda^T f(\xb^{(m)}) \label{eq:LL}
\end{align}
for the spike-data vectors $x^{(m)} \in \{0,1\}^n$, $m = 1, \ldots, M$. 
Any choice of the feature function $f$ defines a specific  maximum entropy model over this $n$-dimensional binary space. 
For the K-pairwise maximum entropy model used in this paper, $f(\xb) \in \{0, 1\}^{n (n+3) /2 + 1}$ is composed of 

\begin{enumerate}
\item $n$ first-order features 
\begin{align} 
f_i(\xb) &= x_i  \nonumber
\end{align}
with corresponding parameters collected in $h$. The $h_i$, $i = 1, \ldots, n$ control single-cell firing rates (in units of bins rather than Hz). 

\item $n (n-1)/2$ second-order features 
\begin{align} 
  f_{ij}(\xb) &= x_i x_j \nonumber
\end{align}
with parameters $J_{ij}$, $j,i = 1, \ldots, n$, $i < j$, controlling pairwise neuronal correlations, and

\item n+1 population-scale features
\begin{align} 
 f_k(\xb) &= 
  \begin{cases}
    1, & \text{if $\sum_i x_i = k$} \\
    0, & \text{otherwise}
  \end{cases}  \nonumber
\end{align}

with parameters $V_k$, $k = 0, \ldots, n$. The vector $V$ controls the overall number of spikes in each temporal bin.
\end{enumerate}

Note that that there is some degeneracy between the parameter vectors $V$ and both $h$ and $J$ ---a global upwards shift of firing rates for example can be achieved both by adding a positive constant $\epsilon$ to each $h_i$, or by adding $\epsilon k$ to each of the $V_k$. 
Similarly, adding a constant $\epsilon$ to every $J_{ij}$ can be balanced by subtracting $\epsilon \frac{k(k-1)}{2}$ from each $V_k$. Since either manipulation of $V$ is zero for $k=0$, fixing $V_{k=0} =0$ is not sufficient for getting rid of this parameter degeneracy. As we never interpreted the parameter-values themselves, but only the fit to data, we made no attempt to add additional constraints.

We can re-write the K-pairwise model into the general maximum entropy form by stacking the feature functions $f_i, f_{ij}$, and $f_k$ into the vector-valued feature function $f$ and doing the same with parameters $h_i$, $J_{ij}$, and $V_k$ to obtain $\lambda = \{h,J,V\}\in \mathbb{R}^{n(n+3)/2+1}$.  The derivative of the log-likelihood with respect to any single parameter $\lambda_l, l = 1, \ldots, n(n+3)/2+1$ is given by
\begin{align}
\frac{\delta}{\delta{}\lambda_l} \sum_{m=1}^M \log P(\xb^{(m)}| \lambda) 
&= \frac{\delta}{\delta{}\lambda_l} \sum_{m=1}^M \left( \lambda^T f(\xb^{(m)}) - \log Z_\lambda \right) \nonumber \\
&= \sum_{m=1}^M \frac{\delta}{\delta{}\lambda_l} \lambda^T f(\xb^{(m)}) - \frac{\delta}
{\delta{}\lambda_l} M \log \sum_\xb \exp\left( \lambda^T f(\xb) \right) \nonumber   \\
&= \sum_{m=1}^M f_l(\xb^{(m)}) - M \frac{\sum_\xb \lambda_l \exp\left(\lambda^T f(\xb)\right)}{\sum_\xb \exp\left(\lambda^T F(\xb)\right)}  \nonumber  \\
&= M \left( \frac{1}{M} \sum_{m=1}^M f_l(\xb^{(m)}) - \text{E}_\lambda[f_l(\xb)] \right)
\label{eq:dLLdlambda}
\end{align}
As can be seen from equation \eqref{eq:dLLdlambda}, the gradient of the log-likelihood vanishes if and only if the data means match the expectations of $f(\xb)$ under the model. 

To deal with data-sets of limited size, we maximised a regularised variant of the log-likelihood, 
\begin{align}
L(h,J,V | \sigma_h, \sigma_J, \Sigma):&=\sum_{m=1}^M \log P(\xb^{(m)} | h,J,V)  - \frac{1}{\sigma_h}\|h\|_1 - \frac{1}{\sigma_J} \|J\|_1 - \frac{1}{2} V^T \Sigma^{-1} V  \label{eq:regLL}
\\
\Sigma &= \left( \sigma_S S + \sigma_\mathds{I} \mathds{I} \right) - \frac{1}{\sigma_S + \sigma_\mathds{I}} S_{0\bullet} {S_{0\bullet}}^T \nonumber \\
S_{k k'}& = \exp\left(-\frac{(k - k')^2}{2 \tau_S^2} \right) \nonumber \\
S_{0 k} &= \sigma_S \exp\left(-\frac{k^2}{2 \tau_S^2}\right). \nonumber
\end{align}
Here, the matrix $\Sigma$ implements a combined ridge and smoothing regression over $V$, with $(n+1)\times(n+1)$ identity matrix $\mathds{I}$ and smoothing matrix $S$ corresponding to a squared-exponential kernel \cite{Rasmussen_Williams_06}. 
We set $V_0=0$ and accounted for this by conditioning on $V_0$ and correspondingly subtracted $S_{0\bullet}(\sigma_S + \sigma_\mathds{I})^{-1} {S_{0\bullet}}^T$ from $\Sigma$. 
We used $\sigma_h=\sigma_J=10^{4}$, $\sigma_S = 10$, $\sigma_\mathds{I} = 400$ and $\tau_S=10$.

To fit maximum entropy models to large neural populations, one needs to 
\begin{enumerate}
\item efficiently approximate the feature moments $\mbox{E}_\lambda[f(\xb)]$ needed for the gradients of both eq. \eqref{eq:LL} and eq. \eqref{eq:regLL}, which for large populations ($n>20$) can not be calculated exactly
\item find efficient methods for updating the parameters $\lambda$. 
\end{enumerate}

We introduce two modifications over previous approaches to fitting maximum entropy models to neural data \cite{Broderick_Dudik_07} to improve computational efficiency:
\begin{enumerate}
\item We used pairwise Gibbs sampling and Rao-Blackwellisation to considerably improve estimation of the second-order feature moments $\mbox{E}_\lambda[f_{ij}(\xb)]$
\item The authors of \cite{Dudik_Phillips_04} described a trick for efficiently updating the parameters in pairwise binary maximum entropy models: If one restricted updates to coordinate-wise updates, then one can calculate the gain from updating a single variable in closed form, which makes it easy to select both the variable to update as well as the step-length in closed form. We show how this trick can be extended to allow a joint update of all the population-count features $V$. In addition, the gain in log-likelihood is linear in the feature-moments, which makes it possible to compute it from a running average over the MCMC sample, and avoids having to store the entire sample in memory at any point. 

\end{enumerate}
We describe our contributions in the sections 
\ref{fitting:pwGibbs_RB} and 
\ref{fitting:bwUpdates}, respectively.

\subsection{Pairwise Gibbs sampling and Rao-Blackwellisation}
\label{fitting:pwGibbs_RB}

Following previous work \cite{Broderick_Dudik_07}, we used MCMC sampling to approximate the expectations of the feature functions $f(\xb)$ under the K-pairwise model with parameters $\lambda$. These expected values $\mbox{E}_\lambda[f(\xb)]$ are required to evaluate the gradients of the (penalised) log-likelihood, as well as the log-likelihood gains resulting from parameter updates. 
As the number of pairwise terms grows quadratically with population size $n$, most of the parameters of the model $P(\xb|\lambda)$ for large $n$ control pairwise moments $\mbox{E}_\lambda[x_i x_j]$.
To make the estimation of these pairwise interactions more efficient, we  implemented a pairwise Gibbs sampler that for each update step of the Markov chain samples two variables $x_i$ and $x_j$, $i \neq j$, $i, j \in 1,\ldots, n$. 
This furthermore allowed us to 'Rao-Blackwellise' the single-cell and pair-wise feature components $f_i(\xb) = x_i$ and $f_{ij}(\xb) = x_i x_j$ \cite{Rao_45, Blackwell_47, Berkson_42}, i.e. to use the conditional probabilities $P( x_i =1 | x_{\sim i}, \lambda)$ and $P( x_i x_j=1 | x_{\sim \{i,j\}}, \lambda)$ for moment estimation, instead of the binary $x_i$ and $x_i x_j$. 

Rao-Blackwellisation provably reduces the variance of the resulting estimators, and empirically resulted in substantially faster convergence of the MCMC-estimated model firing rates $\mbox{E}_\lambda[f_i(\xb)]$, second moments $\mbox{E}_\lambda[f_{ij}(\xb)]$, and thus also of the covariances $\text{cov}_\lambda(\xb_i, \xb_j|\lambda) = \mbox{E}_\lambda[f_{ij}(\xb)] - \mbox{E}_\lambda[f_{i}(\xb)] \mbox{E}_\lambda[f_{j}(\xb)]$ (see supplementary figure \ref{fig_S1}). 
Unlike the binary variables $x_i$, $x_i x_j$ however, the conditional probabilities are real numbers from the interval $(0,1)$ and cannot be stored in memory-efficient sparse matrices.  We thus implemented a running average over conditional probabilities that discards the current chain element immediately after drawing the next one, while keeping track of the quantities 
\begin{align}
\mbox{E}_\lambda[f_{i}(\xb)] &\approx \frac{1}{\tilde{m}} \sum_{m=1}^{\tilde{m}} P( x_i^{(m)} = 1 | x_{\sim \{i\}}^{(m)}, \lambda)  \nonumber \\
\mbox{E}_\lambda[f_{ij}(\xb)] &\approx \frac{1}{\tilde{m}} \sum_{m=1}^{\tilde{m}} P( x_i^{(m)} x_j^{(m)} = 1 | x_{\sim \{i,j\}}^{(m)}, \lambda)  \nonumber
\end{align}

as $\tilde{m}$ increases from $1$ to MCMC sample size $\tilde{M}$. We also kept track of the non-Rao-Blackwellised estimates 
\begin{align}
\mbox{E}_\lambda[f_{k}(\xb)] &\approx \frac{1}{\tilde{m}} \sum_{m=1}^{\tilde{m}} \delta\left(\sum_{i=1}^n x^{(m)}_i, k\right)  \nonumber
\end{align}
for the expectations of the population-level indicator feature functions $\text{}E_\lambda[f_k(\xb)] = P(K=k|\lambda)$, with Kronecker delta function $\delta(x,y) = 1$ if $x=y$, and $\delta(x,y) = 0$ otherwise. 


We quantified the advantage of Rao-Blackwellising the Gibbs sampler with long Markov chains drawn from the K-pairwise maximum entropy model fits to populations of size $n=100$ drawn from the simulated RGC data. For each investigated parameter fit, we ran two chains under different conditions: a first chain for which we Rao-Blackwellised the single-cell and pairwise feature moments, and a second chain for which we did not.
These Markov chains were run for $\tilde{M}=10^6$ sweeps and hence orders of magnitude longer than had occurred for the invidivual parameter updates within this study, which comprised $800$ to $30000$ sweeps, or $3.96 \times 10^6$ to $1.485 \times 10^6$ individual MCMC chain updates at $n=100$. The long sample runs served to give an approximation for the "true" expected values of the target quantities of interest to us: firing rates $\text{E}_\lambda[f_i(\xb)]$, covariances $\text{cov}_\lambda(\xb_i,\xb_j)$ and population spike count distribution $P(K=k|\lambda)$. 

We quantified the speed of convergence of the estimates to the "true" expected feature moments by the normalised MSE between sampler-derived feature moments after any given length $0 < \tilde{m} < \tilde{M}$ of the MCMC chain and the results we got after the full chain length.
After the full $\tilde{M}=10^6$ sweeps, the Rao-Blackwellised and non-Rao-Blackwellised estimates on average differed by $1.7\times 10^{-4} \%$, $0.013 \%$ and $4\times 10^{-6}\%$ normalised MSE for firing rates, covariances and population spike count distributions, respectively. We computed the distance to "truth" for each condition as the normalised MSE to the $\text{E}_\lambda[f(\xb)]$ averaged over both conditions. We obtained MCMC estimates for the feature moments of the K-pairwise maximum entropy models fits to $10$ subsampled populations of $n=100$ neurons each drawn from our retina simulation. Supplementary figure \ref{fig_S1}a displays the results for the two conditions, Rao-Blackwellised vs. non-Rao-Blackwellised, for each of the $10$ investigated fits. 

MSEs of firing rates for single-cell features $\text{E}_\lambda[\xb_i]$ did not benefit from Rao-Blackwellisation. This is expected, as each $x_i$ is sampled $n-1$ times per sweep and thus the moments are already well estimated relative to the second-order features. For covariances $\text{cov}_\lambda(\xb_i,\xb_j)$, normalised MSEs showed clear improvement under Rao-Blackwellisation, visible as an approximately constant offset between the avarages over all $10$ parameter fits in the loglog-domain as seen in figure \ref{fig_S1}b. 
The normalised MSE on average was $3.19$ times higher for non-Rao-Blackwellised (given by the downwards offset of the normalised MSEs of the Rao-Blackwellised estimates). 
The fraction of samples needed from Rao-Blackwellised runs to achieve the same normalised MSE on the pariwise moments than non-Rao-Blackwellised runs (given by the leftward offset of the normalised MSEs of the Rao-Blackwellised) overall was $32.02\%$. 
The fraction ranged from $34.93\%$ at $800$ sweeps to $31.74\%$ at $30000$ sweeps. The ratio of normalised MSEs was similarly stable, being $2.96$ times higher at $800$ sweeps and $3.27$ times higher at $30000$ sweeps for non-Rao-Blackwellised samples than for Rao-Blackwellised ones. 

\begin{figure}[h]
\hspace{-1cm}
\begin{center}
 \includegraphics[width=0.8\textwidth]{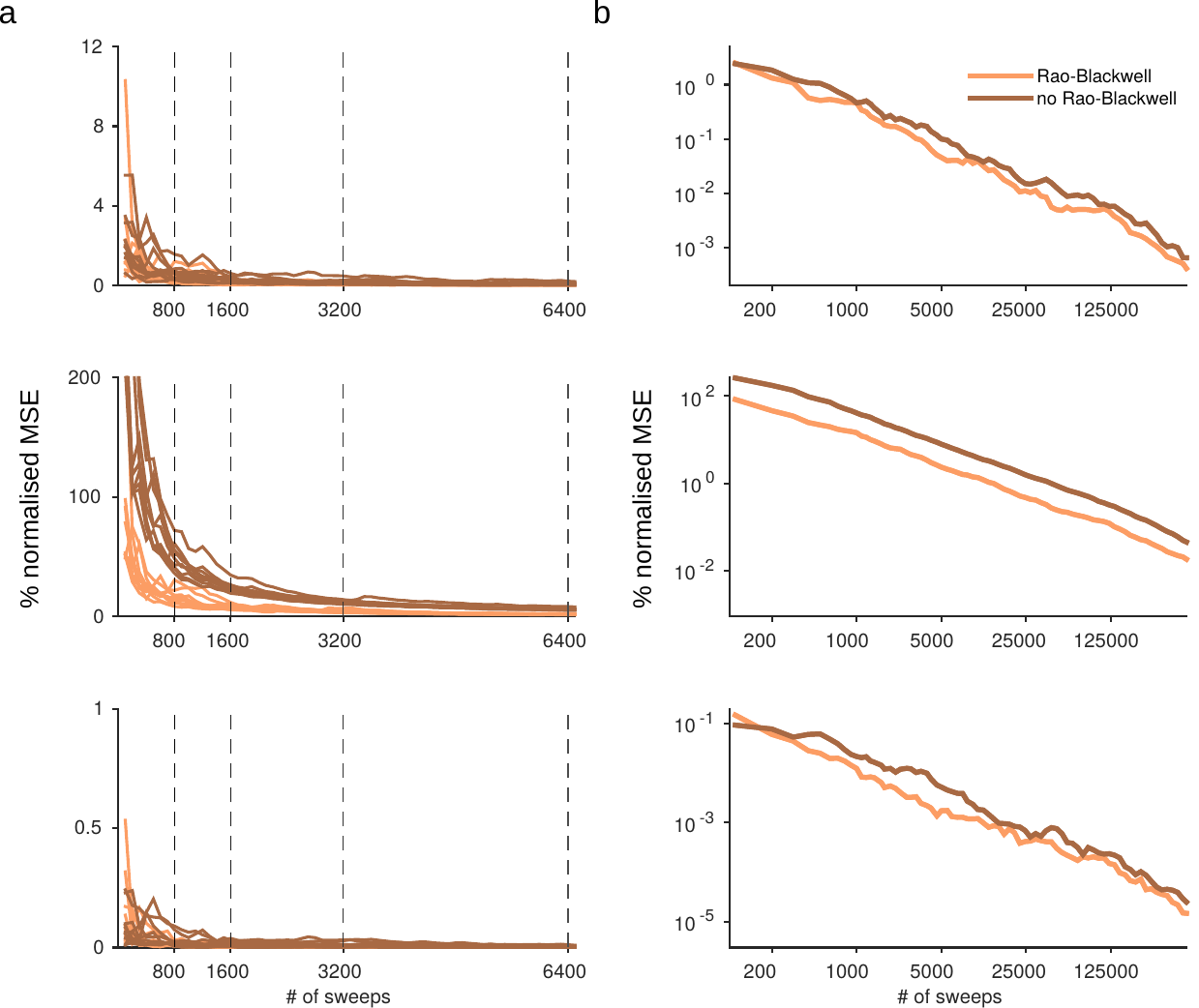} 	
\end{center}
\caption{
{\bf Impact of Rao-Blackwellisation} 
{\bf a)} Comparison of normalised MSE between Rao-Blackwellised and non-Rao-Blackwellised Gibbs sampling, as a function of MCMC chain length, on the $10$ subpopulations of size $n = 100$ used in the paper.   Top: means, i.e. first-order moments $\text{E}_\lambda[\xb_i]$, Center: covariances $\text{cov}_\lambda(\xb_i,\xb_j)$, Bottom: population-spike count features. No Rao-Blackwellization was used for population-spike count features $P(K=k|\lambda)$. Vertical lines and horizontal axis ticks mark Markov chain lengths used for computing the 1st, 1001st 2001st, ... updates of parameter entries $\lambda_l$ during training the K-pairwise models to data.
All MSEs in this figure are computed as errors between estimated firing rates / covariances / $P(K)$ at given chain length versus the average of the estimates obtained after $10^6$ sweeps. 
{\bf b)} Behavior of MSEs for large MCMC chain lengths. 
Traces are averages over the $10$ traces from panel a.
\label{fig_S1} }
\end{figure}

\subsection{Exploiting the structure of the K-pairwise feature functions allows blockwise parameter updates.}
\label{fitting:bwUpdates}

As described in the previous section, we can use MCMC to obtain the expected values of the feature function $\text{E}_\lambda[f(\xb)]$ that are needed to to optimise the model parameters $\lambda$. To find the parameter setting $\hat{\lambda}$ which maximise the log-likelihood over the given data  vectors $\xb^{(m)}$, $m = 1$, \ldots, $M$, we follow  
an iterative update scheme introduced previously \cite{Dudik_Phillips_04}, and extend it to the K-pairwise model. 
The update scheme optimises parameter changes $\lambda^{new} - \lambda^{old}$ relative to a current parameter estimate $\lambda^{old}$, rather than the parameters $\lambda$ directly. 
The benefit of this scheme over standard gradient ascent on the regularised log-ligkelihood as in eq. \eqref{eq:dLLdlambda} is that we can give closed-form solutions for optimal values of a single component $\lambda_l$ when temporarily holding all other components $\lambda_{\sim l}$ fixed. 


Changing the current parameter estimate $\lambda^{old}$ to $\lambda^{new}$ leads to a change in log-likelihood of  
\begin{align}
\Delta{}L(\lambda^{new},\lambda^{old}) &= \frac{1}{M} \sum_{m=1}^M \log P(\xb^{(m)}|\lambda^{new}) - \frac{1}{M} \sum_{m=1}^M \log P(\xb^{(m)}|\lambda^{old}) \nonumber \\ &= (\lambda^{new}-\lambda^{old})^T \left( \frac{1}{M} \sum_{m=1}^M f(\xb^{(m)}) \right) - \mbox{E}_{\lambda^{old}}\left[\exp \left(  (\lambda^{new}-\lambda^{old})^T f(\xb) \right) \right] \label{eq:dLL} 
\end{align}
The only relevant expectations are w.r.t. the data distribution and $P(\xb|\lambda^{old})$, i.e. the current parameter estimate. The term $\mbox{E}_{\lambda^{old}}[\exp \left(  (\lambda^{new}-\lambda^{old})^T f(\xb) \right) ]$ can be simplified when restricting the update vector $\lambda^{new}-\lambda^{old}$ to be non-zero only in selected components.
In the simplest case, only a single component $\lambda_l$ is updated. In this case, the fact that all components of the   K-pairwise feature function $f(\xb)$ are binary, allows to move the exponent out of the expected value, a trick used by \cite{Dudik_Phillips_04}:

 The resulting single-coordinate updates only require the feature moments $\text{E}_{\lambda^{old}}[f_l(\xb)]$:
\begin{align}
\mbox{E}_{\lambda^{old}}[\exp \left(  (\lambda^{new}-\lambda^{old})^T f(\xb) \right) ] &= \text{E}_{\lambda^{old}}[\exp(  (\lambda^{new}_l-\lambda^{old}_l) f_l(\xb)]  \nonumber \\
&= \text{E}_{\lambda_{old}}[1 + (\exp(\lambda^{new}_l-\lambda^{old}_l) -1) f_l(\xb)]  \nonumber \\
&= 1 + (\exp(\lambda^{new}_l-\lambda^{old}_l) -1) \text{E}_{\lambda_{old}}[f_l(\xb)]  \nonumber
\end{align}

Equation \eqref{eq:dLL}  can now be solved analytically for the single free component $\lambda^{new}_l$ that maximises the change in log-likelihood. A closed-form optimal solution is still possible when adding an $l1$-penalty to the log-likelihood \cite{Dudik_Phillips_04}.
We use this $l1$-regularised variant to calculate the possible gain in penalized log-likelihood for each possible update of the single-cell ($h_i$) and pairwise ($J_{ij}$) feature moments $\mbox{E}[x_i]$ and $\mbox{E}[x_i x_j]$, and then perform the update which yield the largest gain.

If we instead allow more than a single component $l$ of the update $\lambda^{new}_l-\lambda^{old}_l$ to be non-zero, we in general would have to deal with the term
\begin{align}
\text{E}_{\lambda_{old}} \left[ \prod_{l \in J} [1 + (\exp(\lambda^{new}_l-\lambda^{old}_l) -1) f_l(\xb) ] \right]  \nonumber
\end{align}
which requires the higher-order moments $\text{E}_{\lambda_{old}} \left[ \prod_{l \in I} f_l(\xb) \right]$ for all $I \subseteq J$ and $J \subseteq \{1, \ldots, n\}$ being the index set of components that are not set to zero.  

The population spike count features $f_k(\xb)$, however, are mutually exclusive (only one of the n+1 features can be non-zero at any time), and therefore we can all parameters of $V$ jointly, and still pull the expectation term outside of the expectation. For the population-spike count features $f_k(\xb)$, hereafter collectively called $f^V(x) \in \{0,1\}^{n+1}$, all such terms of order $||I|| > 1$ are zero due to the sparsity of $f^V(x)$. When restricting the current parameter update of $\lambda$ to only update components corresponding to $V$, we have
\begin{align}
\Delta{L(V^{new},V^{old}}) &= (V^{new}-V^{old})^T \left( \frac{1}{M} \sum_{m=1}^m f^V(\xb^{(m)}) \right) - \mbox{E}_{\lambda^{old}}[\exp(  (V^{new}-V^{old})^T f^V(\xb)) ]  \nonumber
\end{align}
and
\begin{align} 
\mbox{E}_{\lambda^{old}}[\exp \left(  (V^{new}-V^{old})^T f^K(\xb) \right)] 
&= \sum_\xb \exp \left(  (V^{new}-V^{old})^T f^K(\xb) \right) P(\xb|\lambda^{old})  \nonumber \\
&= \sum_{k=0}^{n}\sum_{\xb : \sum_i x_i = k} \exp \left(  (V^{new}_k -V^{old}_k f^K_k(\xb) \right) P(\xb|\lambda^{old}) \nonumber \\
&= \sum_{k=0}^{n} \exp \left(  (V^{new}_k-V^{old}_k) f^K_k(\xb) \right) \sum_{\xb : \sum_i x_i = k}  P(\xb|\lambda^{old}) \nonumber \\
&= \sum_{k=0}^{n} \exp \left(  (V^{new}_k-V^{old}_k) f^K_k(\xb) \right) P(k|\lambda^{old}) \nonumber 
\end{align}

We obtained estimates of the values of $P(k|\lambda^{old}) = \mbox{E}_{\lambda^{old}}[ f_k(\xb) ]$ from the MCMC sample using the indicator functions $f_k(\xb)$, and optimising w.r.t. $V^{new}_k$, $k \in \{1$,\ldots,$n\}$ using gradient-based methods \cite{Schmidt_05}.  

In summary, our update-scheme for maximising the log-likelihood proceeds as follows: For a given parameter vector $\lambda^{old}$, we first estimate the expectation of the feature functions $f_i(\xb)$, $f_{ij}(\xb)$ and $f_k(\xb)$ using a running average over an MCMC sampling and Rao-Blackwellization. We then calculate, for each possible single-neuron parameter $h_i$ and each possibly pairwise term $J_{ij}$ the gain in penalised log-likelihood that we would get from updating it, using methods as described above and derived in \cite{Dudik_Phillips_04}. We additionally compute the gain in penalised log-likelihood that would result from optimising all $n$ of the free $V$ parameters jointly, using a convex optimization. Finally, we choose the update that brings the largest gain, and either update a single $h_i$, a single $J_{ij}$, or all $V$ parameters. Subsequently, we again estimate the new feature functions using MCMC sampling given the current estimate of $\lambda^{old} \leftarrow \lambda^{new}$ before we update again. We initialised the algorithm assuming independent neurons (i.e. setting each $h_i$ using the firing rate of each neuron, and leaving $J$ and $V$ zero). The algorithm then typically first updated all $V$ parameters, before proceeding to jump between different $J$, $h$ and $V$ updates.

\begin{figure}[h]
\hspace{-1cm}
 \begin{center}
  \includegraphics[width=0.8\textwidth]{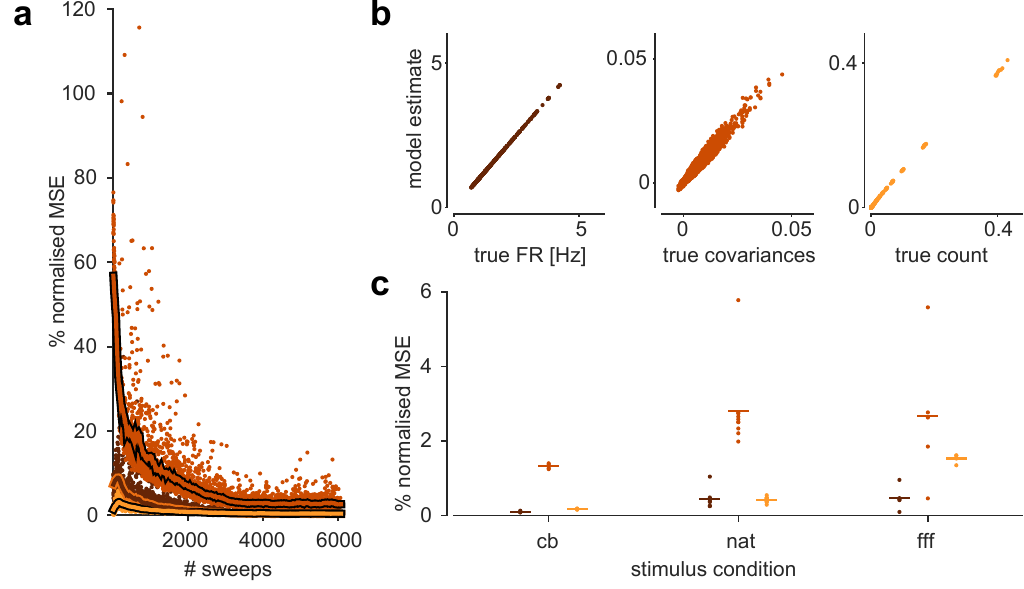} 	
 \end{center}
\caption{
{\bf Quality of fits for K-pairwise maximum entropy model across multiple populations and stimulus conditions} 
{\bf a)} Normalised MSEs for firing rates, covariances and $P(K)$ during parameter learning. Error values collapses across $10$ subpopulations at $n=100$, fit to simulated activity in response to natural images, one point for each displayed iteration and each subpopulation.  Lines are moving averages (smoothing kernel width = 150 param. updates).  
 {\bf b)} Quality of fit after parameter learning. Data vs. model estimates for firing rates, covariances and $P(K)$, collapsed over all $10$   subpoplations with size $n = 100$. {\bf c)} Quality of fit for different stimulus types. Normalised MSEs after maximum entropy model fitting shown for $10$ subpopulations for natural images (nat) and $5$ subpopulations each for checkerboard (cb) and full-field flicker (fff). All subpopulations of size $n = 100$. Vertical bars give averages. Colours as in {\bf a), b)}. 
\label{fig_S2} }
\end{figure}  


\section{Supplementary Text: Specific heat in simple models}
\label{flat:divergence_proof}

We refer to a maxmimum entropy model as 'flat' if it is fully specified by the population spike count distribution $P(\sum_{i=1}^n x_i = k)$, i.e. the model class studied in \cite{Macke_Opper_11, Tkacik_Marre_13, Amari_Nakahara_03}.  In this model class, all neurons have the same firing rate $\mu$ and pairwise correlation $\rho$. 
As neuron identities become interchangeable, all $n \choose k$ possible patterns $\xb$ with $\sum_{i=1}^n = k$ are assigned the same probability $P(k) = P(\xb) {n \choose k} $. In flat models, all relevant population properties can be computed from summing over $n+1$ different spike counts, and one never has to (explicitly) sum over the entire $2^n$ possible spike patterns.



\subsection{A non-critical special case: Independent neurons}

A special case of a flat model is an independent model in which all neurons have the same firing rates and zero correlations. Assuming independent spiking for each of the $n$ neurons and a shared probability $q  \in [0,1]$ to fire in a time bin, the distribution of population spike counts $K = \sum_{i=1}^n x_i $ is given by a binomial distribution,
\begin{align}
P(\xb | q)  &=  q^k (1-q)^{n-k}  \nonumber\\
P(k| q) &= {n \choose k} q^k (1-q)^{n-k}  \nonumber
\end{align}


To compute specific heat capacities for the underlying neural population of size $n$, we can rewrite the binomial distribution in maximum entropy form 
\begin{align}
P(\xb | V) &= \frac{1}{Z(V)} \exp\left( V_k \right)  \nonumber \\
P(k | V) &= \frac{1}{Z(V)} {n \choose k} \exp\left(  V_k  \right)  \nonumber
\end{align}
Re-introducing parameters $V_k$, $k =  0, \ldots, n$, we find
\begin{align}
V_k &= \log P(k|q) - \log {n \choose k} + \log Z(V)  \nonumber \\
&= k \log(q) + (n-k) \log(1-q) ) \nonumber 
\end{align}
and for the heat capacity, we get
\begin{align}
\mbox{Var}[\log P(x|V)] &= \mbox{Var}[k \log(q) + (n - k) \log(1-q)]  \nonumber \\                                                        &= (\log(q) - \log(1-q))^2 \hspace{0.1cm} \mbox{Var}[k] \nonumber 
\end{align}
The binomial variance is $\mbox{Var}[k] = n q (1-q)$.
We plug this in and see that at unit temperature $T=1$, the specific heat is given by
\begin{align}
c(T=1) =  \frac{1}{n T^2} \mbox{Var}[\log P(\xb|V)] =  q (1-q) (\log(q) - \log(1-q))^2 \label{eq:cT1_independent}
\end{align}
which is independent of population size $n$.

When explicitly introducing temperatures other than $T=1$, we add a factor $\frac{1}{T} = \beta$ that scales the parameters $V$ and renormalise, yielding
\begin{align}
 P(k | V, T) = \frac{1}{Z(\beta V)} {n \choose k} \exp(\beta  V_k) \nonumber
\end{align}
where $V_k$, $k = 0$, ..., $n$ is defined w.r.t. $q$ as above. This is the same functional form as was given for the binomial distribution at $T=1$, with only parameters $V$ being replaced by $\beta V$. We can also go back to the standard binomial parametrisation with $q_\beta = \frac{q^\beta}{q^{\beta} + (1-q)^\beta}$ and obtain
\begin{align}
P(k| V, T) = {n \choose k} q_{\beta}^k (1-q_\beta)^{(n - k)}  \nonumber
\end{align}
Changing the temperature $T = \frac{1}{\beta}$ retains the binomial form of the population model, and we can generalise the expression for the specific heat \eqref{eq:cT1_independent} of the independent flat model for any temperature $T$ to be
\begin{align}
c(T) =  \frac{1}{T^2} q_\beta (1-q_\beta) (\log(q_\beta) - \log(1-q_\beta))^2 \nonumber
\end{align}
which again is independent of the population size $n$. The independent flat model is a case that does not show divergent specific heat, and for which the peak of the heat is not necessarily at unit temperature. Next, we will derive why this makes the binomial model one of only two non-critical special cases.

\subsection{Aside: Asymptotic entropy in flat models}

To calculate the variance of log-probabilities, we first need the mean log-probability, i.e. the (negative) entropy.

\paragraph{Entropy:}

Recalling that $P(k) = P(\xb) {n \choose k} $, the entropy of the flat model for general $P(k)$ can be written as
\begin{align}
H_n& = - \sum_x P(\xb) \log P(\xb)  \nonumber \\
&= -\sum_k \sum_{\xb: \sum_i x_i = k} P(\xb) \log P(\xb) \nonumber \\
&= -\sum_k   P(k)  \left( \log P(k)-\log{n \choose k}\right) \nonumber 
\end{align}

Thus, the entropy of a flat model is
\begin{align}
H_n &= -\sum_k   P(k)  \left( \log P(k) -\log{n \choose k}\right)   \nonumber
\end{align}

\paragraph{Asymptotic entropy:}
We assume that $P(k)$ has a limiting distribution $f(r)$, where $r \in [0,1]$ is the probability density of a proportion of $r$ neurons spiking simultaneously. 
Therefore, for large $n$
\begin{align}
H_n &= -\sum_k   P(k)  \left( \log P(k) -\log{n \choose k}\right)  \nonumber \\
&\approx -\sum_k \frac{1}{n}f\left(\frac{k}{n}\right)  \left( \log P(k) -\log{n \choose k}\right) \nonumber  \\
& \approx -\int_{0}^1  f(r) \left( \log \frac{f(r)}{n} -\log{n \choose nr}\right) dr \nonumber \\
& = -\log(n) \int_{0}^1  f(r) \log f(r) dr + n \int_{0}^1  f(r) \eta(r) dr \nonumber 
\end{align}

Here, we used the fact that, for large $n$,
\begin{align}
\log{n \choose nr} \approx n \left( -r \log r- (1-r) \log(1-r) \right) =: n \eta(r) 
\end{align}

As the first term only grows with $\log(n)$, and the second with $n$, we get that the entropy of a flat model,  for large $n$, is given by
\begin{align}
H_n=  n \int_{0}^1  f(r) \eta(r) dr=: n h
\end{align}

\subsection{Asymptotic specific heat in flat models at unit temperature}

Next, we calculate the specific heat, first exactly and then for large $n$, and finally for weakly correlated models:  

First, the specific heat is given by 
\begin{align}
c(T=1)&=\frac{1}{n}\mbox{Var}[\log P(\xb)] = \frac{1}{n} \sum_x P(\xb) \left(\log P(\xb)- \text{E}[ \log P(\xb) ] \right)^2  \nonumber \\
&=  \frac{1}{n}\sum_k P(k) \left( \log P(k)- \log{n \choose k} - \text{E}[ \log P(\xb)]\right)^2 \nonumber
\end{align}

Using $\text{E}[\log P(\xb)] =-H_n$, we get that
\begin{align}
c(T=1)&=\frac{1}{n} \sum_k P(k) \left( \log P(k) - \log{n \choose k} + H_n \right)^2 \text{or}  \nonumber \\
&= \frac{1}{n} \sum_k P(k) \left( \log P(k) - \log{n \choose k}\right)  - \frac{1}{n} H_n^2 \nonumber
\end{align}

For large $n$, we have that $P(k)\approx  \frac{1}{n} f\left(\frac{k}{n}\right)$. We get that
\begin{align}
c(T=1)&= \frac{1}{n}\sum_k P(k) \left(\log P(k)- \log{n \choose k}+ H_n\right)^2  \nonumber \\
&\approx \frac{1}{n} \sum_k \frac{1}{n} f\left(\frac{k}{n}\right) \left(\log\left(\frac{1}{n} f\left(\frac{k}{n}\right)\right)- \log{n \choose k}+ H_n\right)^2 \nonumber \\
&\approx \frac{1}{n} \int_0^1  f\left(r\right) \left(\log f\left(r\right) - \log n - \log{n \choose nr}+ H_n\right)^2  dr \nonumber \\
&\approx \frac{1}{n} \int_0^1  f\left(r\right) \left(\log f\left(r\right) - \log n - n \eta(r) +n h_n\right)^2  dr  \nonumber \\
&= \frac{1}{n}\int_0^1  f\left(r\right) \left( \left( \log f\left(r\right) - \log n \right)^2+  n^2 \left( \eta(r) - h_N\right)^2+ 2n \left( \log f\left(r\right)- \log n \right) \left(h_n-\eta(r) \right) \right)  dr  \nonumber  \\
&= \frac{1}{n}\int_0^1  f\left( r \right) \left( \log^2 f\left( r \right) + \log f\left( r \right) \left( 2n  \left(h_n-\eta(r)\right)  -2 \log n \right) \right) dr \nonumber  \nonumber \\
& + \frac{1}{n}\int_0^1 f\left( r \right) \left(   
\log^2 n - n^2 \left( h_n - \eta(r) \right)    
\right) dr \nonumber
\end{align}

For large $n$, this integral is dominated by the term in $n^2$, and thus the specific heat is asymptotically given by 

\begin{align}
c(T=1)= n \int_0^1 f(r) \left(\eta(r)- h \right)^2 dr 
\label{eq:cT1_startingPoint}
\end{align}

\begin{figure}[h]
\hspace{-1cm}
\begin{center}
 \includegraphics[width=0.6\textwidth]{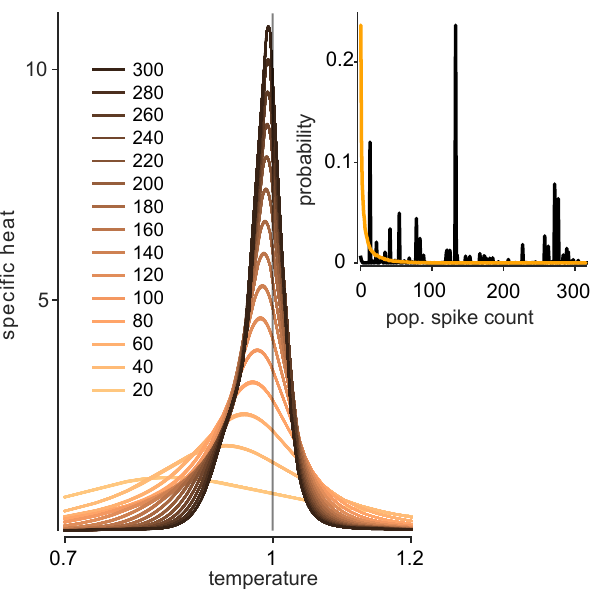} 	
\end{center}
\caption{
{\bf Diverging specific heat for a non-natural spike-count distribution}
The values of the population spike count distribution $P(K)$ obtained from the retinal simulation with $N=316$ in response to natural image stimulation (orange, inset) were shuffled (black trace, inset) across $K$, to yield a 'pathological' $P(K)$.  
We simulated data for this $P(K)$ from a flat model, and subsampled subpopulations of size $n = 20, \ldots, 300$. 
The specific heat traces computed from this data also diverges and has a peak at unit temperature \label{fig_S3}
}
\end{figure}

Therefore, in general, the specific heat grows linearly, and hence diverges (see Fig \ref{fig_S3}). 
The only exception to this are models for which  $\eta(r)- h_n = 0$ for almost all $r$. 
This happens if $f(r)$ is a delta-distribution, $f(r)= \delta(r-\mu)$, in which case $h_n= \eta(\mu)$ and therefore the integral vanishes. 
This occurs whenever the pairwise correlations do not grow proportionally with $n^2$, as then the variance of the population spike count collapses in the limit. 
One such special case is the binomial distribution over $k$, as already demonstrated above using a more direct approach. 
There is a second special case, namely if $f(r)$ is a combination of two $\delta$-peaks at $\mu$ and $1-\mu$ (See \cite{Macke_Opper_11} for details)-- this special case corresponds to a flat Ising model. 

\subsection{In flat models, specific heat does not diverge for temperatures which are not equal to $1$:} 

Above we showed that at unit temperature, the specific heat for flat models (almost) always diverges. 
Now, we show that this is NOT true for any other temperature. 
This explains that, for any $f(r)$, we will find that the unit temperature is 'special'.

First, we calculate the spike-count distribution at any inverse temperature $\beta$:
\begin{align}
P_\beta(\xb) &= \frac{1}{Z_\beta} P(\xb)^\beta \nonumber \\
P_\beta(k) &= \frac{1}{Z_\beta} {n \choose k}^{1-\beta} P(k)^\beta  \nonumber
\end{align} 

For large $n$,
\begin{align}
f_\beta(r) &\approx n P_\beta(r n)  \nonumber \\
&= \frac{n}{Z_\beta} {n \choose nr}^{1-\beta} P(r n)^\beta \nonumber \\
&\approx \frac{n}{Z_\beta} \exp\left(n(1-\beta) \eta(r) \right) P^\beta(r n) \nonumber
\end{align}

For large populations, this expression is dominated by the exponential term $\exp\left(n(1-\beta) \eta(r) \right)$. 
For $\beta<1$, the exponential term is in turn dominated by the mode of $\eta(r)$, which is at $r=\frac{1}{2}$. 
Thus, for $\beta<1$, $f_\beta(r)= \delta(r-\frac{1}{2})$, a delta-peak at $r=\frac{1}{2}$. 

Conversely, for $\beta>1$, the argument of the exponential has its peaks at $r=0$ and $r=1$, and therefore $f_\beta(r)= \frac{1}{2}\delta(r-1)+ \frac{1}{2}\delta(r-0)$. 
In this case, we also have that the integral in the specific heat vanishes, and that the specific heat does not diverge.

\section{Specific heat divergence rate in flat models as function of correlation strength}
\label{flat:corrs_and_criticality}

In the next two sections, we will derive analytic expressions to predict the specific heat divergence rate in flat models as a function of the correlation strength within the population. Starting out from eq. \eqref{eq:cT1_startingPoint}, we will use two different approximations to $f(r)$ that will each yield results that allow us to better understand the behavior of the specific heat at unit temperature $c(T=1)$ in flat models.

\subsection{Asymptotic entropy and specific heat in weakly correlated flat models:}
\label{flat:corrs_and_criticality_gauss}

Next, we examine entropy and specific heat in models with weak correlations. 
If the model is weakly correlated and its mode is not at $0$ or $1$ we can assume it to be approximately Gaussian with mean $\mu$ and variance $\sigma^2$,
\begin{align}
f(r) = \frac{1}{Z} \exp\left(-\frac{1}{2\sigma^2} \left( r-\mu   \right)^2 \right)  \nonumber. 
\end{align}

We first calculate the entropy: We expand $\eta(r)$ to second order around $\mu$, 
\begin{align}
\eta(r)&= \eta(\mu+\delta)= \eta(\mu)+ \eta'(\mu) \delta+ \frac{\delta^2}{2} \eta''(\mu)+ ..., \mbox{where } \nonumber \\
\eta'(r)&=\log\left(\frac{1-r}{r}\right) \nonumber \\
\eta''(r)&= \frac{-1}{r(1-r)}, \mbox{ so} \nonumber \\
\eta(\mu+\delta)& = \eta(\mu) + \delta \log\left(\frac{1-\mu}{\mu}\right) -\frac{\delta^2}{2\mu(1-\mu)}+ ... \nonumber \\
&=: \alpha + \delta \beta+ \delta^2 \gamma \nonumber
\end{align}

Thus, the asymptotic entropy-rate is given by
\begin{align}
h&= \int f(r) \eta(r) dr  \nonumber \\
&= \alpha+ 0\beta+\gamma \sigma^2 \nonumber \\
&= \eta(\mu)- \frac{1}{2\mu(1-\mu)}\sigma^2 \nonumber
\end{align}

We further investigate the variance, again neglecting all terms which are of higher order than $2$, obtaining
\begin{align}
\left(\eta(\mu+\delta)-h \right)^2& = \left((\alpha-h)+ \beta\delta+ \gamma\delta^2 \right)^2  \nonumber \\
&= (\alpha-h)^2+ \delta^2\beta^2+ 2(\alpha-h)\beta\delta+ 2(\alpha-h)\gamma\delta^2+ 2(\alpha-h)\gamma\delta^2+ \ldots \nonumber \\
&= (\alpha-h)^2+ \delta\left(2(\alpha-h)\beta \right)+ \delta^2\left(\beta^2+ 2(\alpha-h)\gamma \right) + \ldots \nonumber
\end{align}

Integrating this expression over $f(r)$, and dropping all terms in $\sigma$ which are of order higher than $2$, we get
\begin{align}
\int f(r) (\eta(r)- h)^2 &= (\alpha-h)^2 +\sigma^2\left(\beta^2+ 2(\alpha-h)\gamma \right)  \nonumber \\
&=\frac{\sigma^4}{\mu^2(1-\mu)^2}+ \sigma^2\left(\log^2\left(\frac{1-\mu}{\mu}\right)-\frac{\sigma^2}{\mu^2(1-\mu)^2}\right) \nonumber\\
&\approx \sigma^2\log^2\left(\frac{1-\mu}{\mu} \right) \nonumber
\end{align}

In summary, we arrive at
\begin{align}
c(T=1)&=n \sigma^2\log^2\left(\frac{1-\mu}{\mu} \right) \text{ and, for small $\mu$,}  \nonumber \\
c(T=1)&=n \sigma^2 \log^2(\mu)  \nonumber
\end{align}

In other words, a population of a given size $n$ at fixed firing rate $\mu$ that has a high specific heat is simply a population which is very correlated. 
Inspecting the equations above, we see that the final results do not critically depend on the Gaussian assumption---the only requirement for the calculation to be accurate is that the distribution is reasonably peaked around its mean. 

\subsection{Asymptotic specific heat in the beta-binomial population model}
\label{flat:corrs_and_criticality_betabin}

\begin{figure}[h]
\hspace{-1cm}
\begin{center}
 \includegraphics[width=0.8\textwidth]{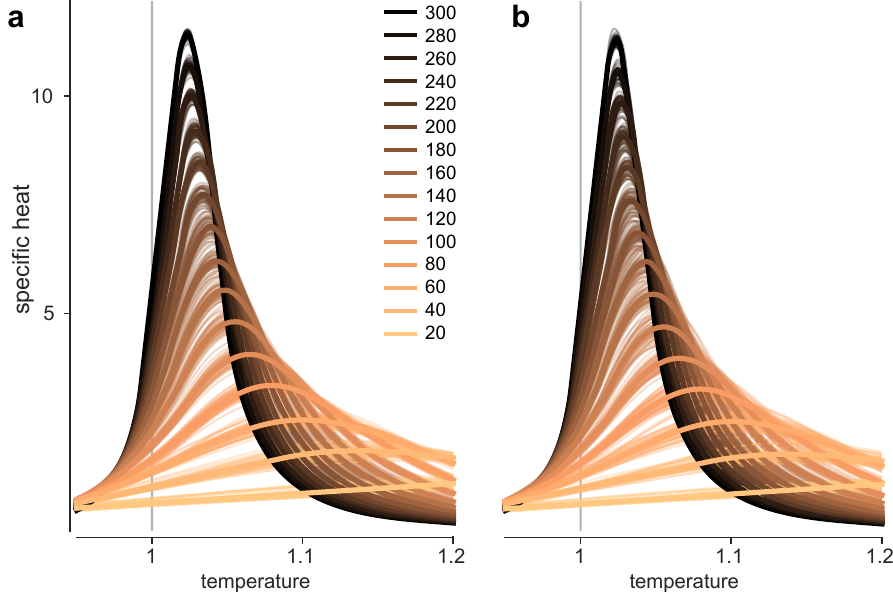} 	
\end{center}
\caption{
{\bf (No) Influence of beta-binomial approximation on heat capacity} 
Specific heat capacities computed from population spike count distributions $P(K)$. Spike count distributions for population sizes $n = 20, \ldots, 300$ were obtained from $50$ uniformly drawn subpopulations each. 
Simulated retinal activity was taken from of the  retina simulation with in total $N=316$ RGCs that responded to natural image stimulation. 
Resulting specific heat traces computed from beta-binomial approximations to the spike count distributions (left) and from raw $P(K)$ (right) do not display visible differences. 
\label{fig_S4}
}
\end{figure}

For the beta-binomial model, we assume  $f(r)$ to be given by a beta distribution, i.e. 
\begin{align}
f(r) =\frac{1}{B(\alpha,\beta)}r^{\alpha-1} (1-r)^{\beta-1}. \nonumber 
\end{align} 
Such $f(r)$ arise for large populations when the population spike count $k$ is described by a beta-binomial distribution, and the choice for the beta distribution as a model for $f(r)$ was motivated by the successful application of beta-binomial models $P(k|\alpha, \beta)$ to our simulated RGC activity (see Fig. \ref{fig_S4}). 

For beta-distributed $r$, we have 
\begin{align}
\mbox{E}[r]&= \frac{\alpha}{\alpha+\beta}, \nonumber \\ 
\mbox{Var}[r]&=\frac{\alpha\beta}{(\alpha+\beta)^2(\alpha+\beta+1)}, \nonumber \\
\mbox{E}[\log r]&= \gamma(\alpha)-\gamma(\alpha+\beta), \nonumber
\end{align}
where $\gamma$ denotes the digamma function.


The entropy can be calculated using known results on the expectation of the log,
\begin{align}
h=  \gamma(\alpha+\beta+1)- \frac{\alpha}{\alpha + \beta} \gamma(\alpha+1)- \frac{\beta}{\alpha + \beta} \gamma(\beta+1)  \nonumber
\end{align}
For the specific heat at unit temperature according to equation \eqref{eq:cT1_startingPoint}, we however also require the expected values $$\mbox{E}[r^2  \log^2 r ], \mbox{E}[(1-r)^2 \log^2(1-r)], \mbox{E}[r (1-r) \log r \log(1-r)]$$ 
i.e. 
\begin{align}
\mbox{E}[r^k (1-r)^l \log^m r \log^n(1-r)] = \int_0^1{  f(r) \{ r^k (1-r)^l \log^m r \log^n(1-r) \} dr} \label{eq:LeibnizIdea}
\end{align}
under beta-binomial distribution $f(r)$, where $k, l, m, n \in \{0,1,2\}$.

We begin the derivation of these terms by observing that 
\begin{align}
u{(m,n)}(r,\alpha+k, \beta+l) &= \log(r)^m r^{(\alpha+k-1)} \log(1-r)^n (1-r)^{(\beta+l-1)} \nonumber \\
\frac{\delta}{\delta{\alpha}} u_{(m,n)}(r,\alpha+k, \beta+l) &= \log(r)^{m+1} r^{(\alpha+k-1)} \log(1-r)^n (1-r)^{(\beta+l-1)} \nonumber \\ &= u_{(m+1,n)}(r,\alpha+k, \beta+l) \nonumber \\
\frac{\delta}{\delta{\beta}} u_{(m,n)}(r,\alpha+k, \beta+l) &= \log(r)^m r^{(\alpha+k-1)} \log(1-r)^{n+1} (1-r)^{(\beta+l-1)} \nonumber \\ &= u_{(m,n+1)}(r,\alpha+k, \beta+l) \nonumber 
\end{align}
for any $k,l \in \mathbb{N}$.  Note that the exponents $k,l$ are readily absorbed into new effective beta distribution parameters $\alpha' = \alpha + k$, $\beta' = \beta + l$. 

The triplets ($u_{(m,n)}$, $u_{(m+1,n)}$,$u_{(m,n+1)})$ for any $m,n \in \mathbb{N}$ recursively express the integrands of \eqref{eq:LeibnizIdea} as continuous derivatives, which allows us to repeatedly apply Leibniz' rule to the integral. 
We first deal with $\mbox{E}[r^k \log^m r]$, where $m=k=2$, $n=l=0$, $\alpha'=\alpha+2$, $\beta'=\beta$, which is the first of the three expected values we need to compute the specific heat at unit temperature:
\begin{align}
\mbox{Beta}(\alpha, \beta) \mbox{E}[r^2 \log^2 r]  &= \int_0^1{ r^{\alpha-1} (1-r)^{\beta-1} \log^2(r) r^2 dr} \nonumber \\
   &= \int_0^1{  \frac{\delta^2}{\delta{\alpha}^2} \{ r^{\alpha+1} (1-r)^{\beta-1} \} dr} \nonumber \\
   &= \int_0^1{  \frac{\delta}{\delta{\alpha}} \{ \frac{\delta}{\delta{\alpha}} \{ r^{\alpha+1} (1-r)^{\beta-1} \} \} dr} \nonumber \\
   &= \frac{\delta}{\delta{\alpha}} \int_0^1{   \frac{\delta}{\delta{\alpha}} \{ r^{\alpha+1} (1-r)^{\beta-1} \} dr} \nonumber \\ 
   &= \frac{\delta^2}{\delta{\alpha}^2} \int_0^1{   r^{\alpha+1} (1-r)^{\beta-1} dr} \nonumber \\ 
   &= \frac{\delta^2}{\delta{\alpha}^2}  \mbox{Beta}(\alpha+2, \beta) \nonumber 
\end{align}

The first two derivatives of $\mbox{Beta}(\alpha',\beta')$ w.r.t. $\alpha$ are given by
\begin{align}
 \frac{\delta}{\delta{\alpha}} \mbox{Beta}(\alpha',\beta') &= \mbox{Beta}(\alpha',\beta')  ( \psi_0(\alpha') - \psi_0(\alpha' + \beta')) \text{ and}\nonumber \\
 \frac{\delta^2}{\delta{\alpha^2}} \mbox{Beta}(\alpha',\beta')
&= \mbox{Beta}(\alpha',\beta') \left( ( \psi_0(\alpha') - \psi_0(\alpha' + \beta'))^2  + \psi_1(\alpha') - \psi_1(\alpha'+\beta') \right). \nonumber
\end{align}

We obtain the $m$-th derivative also for $m>2$ using an iterative rule.
The beta-binomial normaliser $\mbox{Beta}(\alpha',\beta')$ furthermore cancels out with the denominator $\mbox{Beta}(\alpha, \beta)$ of the original beta distribution through
\begin{align}
\mbox{Beta}(\alpha + k , \beta + l ) = \frac{ \prod_{i=0}^{k-1} (\alpha + i) \prod_{j=0}^{l-1} (\beta + j) }{\prod_{i=0}^{k+l-1} (\alpha + \beta+ i)} \mbox{Beta}(\alpha, \beta) \nonumber 
\end{align} 
Combining the previous results gives
\begin{align}
 \mbox{E}[r^2 \log^2 r] 
 &= \frac{1}{\mbox{Beta}(\alpha,\beta)} \int_0^1{ r^{\alpha-1} (1-r)^{\beta-1} log^2(r) r^2 dr} \label{eq:betaProof_r2log2r} \\
 &=  \frac{1}{\mbox{Beta}(\alpha,\beta)} \frac{\delta^2}{\delta{\alpha}^2}  \mbox{Beta}(\alpha+2, \beta) \nonumber \\
 &= \frac{\alpha(\alpha+1)}{(\alpha+\beta)(\alpha+\beta+1)} \frac{1}{\mbox{Beta}(\alpha+2,\beta)} \frac{\delta^2}{\delta{\alpha}^2}  \mbox{Beta}(\alpha+2, \beta) \nonumber \\
 &= \frac{\alpha(\alpha+1) \left( ( \psi_0(\alpha +2 ) - \psi_0(\alpha + \beta +2))^2  + \psi_1(\alpha + 2) - \psi_1(\alpha+\beta +2) \right)}{(\alpha+\beta)(\alpha+\beta+1)}.  \nonumber 
\end{align}
For $m=2$, $k=1$, $n, l = 0$ the result
$$E[r \log^2 r] = \frac{\alpha}{\alpha + \beta} [ (\psi_0(\alpha+1) - \psi_0(\alpha+\beta+1))^2 + \psi_1(\alpha +1) - \psi_1(\alpha+\beta+1)]$$
is identical to the one from \cite{Archer_Park_14} in the appendix A.3, eq. (28). \\

We have $\mbox{Beta}(\alpha,\beta) = \mbox{Beta}(\beta,\alpha)$, i.e. the above equations hold symmetrically for $\alpha$ and $\beta$ interchanged, and $n,l$ instead of $m,k$. 
This gives us the second required term to compute the specific heat at unit temperature, 
\begin{align}
 &\mbox{E}[(1-r)^2 \log^2 (1-r)]  \label{eq:betaProof_1mr2log21mr}\\
 &= \frac{\beta(\beta+1) \left( ( \psi_0(\beta +2 ) - \psi_0(\alpha + \beta+2))^2  + \psi_1(\beta + 2) - \psi_1(\alpha+\beta +2) \right)}{(\alpha+\beta)(\alpha+\beta+1)}  \nonumber 
\end{align}

Including derivatives w.r.t. both $\alpha$ and $\beta$, we more generally arrive at
\begin{align}
 \mbox{E}[\log(r)^m r^k \log(1-r)^n (1-r)^l]  
 &= \frac{ \prod_{i=0}^{k-1} (\alpha + i) \prod_{j=0}^{l-1} (\beta +j) }{\prod_{i=0}^{k+l-1} (\alpha + \beta+ i)}  g_{(m,n)}(\alpha + k, \beta +l). \nonumber 
\end{align}

We get recursive formulas for $g_{(m,n)}$, starting at $g_{(0,0)}(\alpha, \beta) = 1$: 
\begin{align}
g_{(m+1,n)}( \alpha, \beta) &= \left( \psi_0(\alpha) - \psi_0(\alpha + \beta) \right) g_{(m,n)}(\alpha, \beta) + \frac{\delta}{\delta{\alpha}} g_{(m,n)}(\alpha + \beta)  \nonumber  \\
g_{(m,n+1)}(\alpha, \beta) &= \left( \psi_0(\beta) - \psi_0(\alpha + \beta) \right) g_{(m,n)}(\alpha, \beta) + \frac{\delta}{\delta{\beta}} g_{(m,n)}(\alpha + \beta). \nonumber 
\end{align}

To compute $c(T=1)$, we still require the case of $m=k=n=l=1$ given by 
\begin{align}
 \mbox{E}[r(1-r) \log(r)\log(1-r)  ]  &= \frac{ \alpha\beta}{(\alpha + \beta)(\alpha + \beta+ 1)} g_{(1,1)}(\alpha + 1, \beta + 1)  \label{eq:betaProof_rrm1logrlogrm1}
\end{align}
with 
\begin{align}
g_{(1,1)}(\alpha + k, \beta + l) &=
\psi_0(\alpha + k) \psi_0(\beta + l) - \psi_0(\alpha + \beta + k + l) \left( \psi_0(\alpha + k) + \psi_0(\beta + l) \right)  \nonumber \\
&+ \psi_0(\alpha + \beta + k + l)^2 - \psi_1(\alpha + \beta + k + l) \label{eq:betaProof_rrm1logrlogrm1_g}.
\end{align}

Combining the results of equations
\eqref{eq:betaProof_r2log2r},
\eqref{eq:betaProof_1mr2log21mr},
\eqref{eq:betaProof_rrm1logrlogrm1}, \eqref{eq:betaProof_rrm1logrlogrm1_g} with eq. \eqref{eq:cT1_startingPoint}, we arrive at 
\begin{align}
\frac{c(T=1)}{n} &= \int_0^1 f(r) \left(\eta(r)- h \right)^2 dr \nonumber \\
&= \frac{\alpha (\alpha +1) \psi_1(\alpha+1) + \beta (\beta +1) \psi_1(\beta+1)}{ (\alpha + \beta) (\alpha + \beta +1 ) } \nonumber \\
  &+ \frac{\alpha \beta \left( \psi_0(\alpha +1) - \psi_0(\beta +1) \right)^2 }{ (\alpha + \beta)^2 (\alpha + \beta +1 ) } - \psi_1(\alpha + \beta +1 ).      \nonumber
\end{align}

\end{document}